\documentclass[11pt,a4paper]{article}
\pdfoutput=1

\usepackage{amsmath, amssymb}
\usepackage[T1]{fontenc}

\usepackage{jheppub}
\usepackage{psfrag}
\usepackage{slashed}
\usepackage{cancel}
\usepackage{lscape}

\usepackage{caption}
\usepackage{subcaption}
\usepackage{array}
\usepackage{graphicx}
\usepackage{subcaption}
\usepackage{multirow}
\usepackage{float}
\usepackage[utf8]{inputenc}

\newcommand\njet{\textsc{Njet}}
\renewcommand\to{\rightarrow}

\DeclareMathOperator{\EX}{\mathbb{E}}

\preprint{{IPPP/20/5}}

\title{Using neural networks for efficient evaluation of high multiplicity scattering amplitudes}

\author[a]{Simon Badger,}
\author[a,b]{Joseph Bullock}
\affiliation[a]{
Institute for Particle Physics Phenomenology, Department of Physics, Durham University, Durham DH1
3LE, United Kingdom%
}
\affiliation[b]{
Institute for Data Science, Durham University, Durham DH1
3LE, United Kingdom%
}
\emailAdd{simon.d.badger@durham.ac.uk, j.p.bullock@durham.ac.uk}

\abstract{Precision theoretical predictions for high multiplicity scattering rely on
the evaluation of increasingly complicated scattering amplitudes which come with an extremely high CPU cost.
For state-of-the-art processes this can cause technical bottlenecks in the production of fully
differential distributions. In this article we explore the possibility of using neural networks to approximate
multi-variable scattering amplitudes and provide efficient inputs for Monte Carlo integration. We focus on
QCD corrections to $e^+e^-\to$ jets up to one-loop and up to five jets. We demonstrate reliable interpolation when a
series of networks are trained to amplitudes that have been divided into sectors defined by their infrared singularity structure.
Complete simulations for one-loop distributions show speed improvements of at least an order of magnitude over a standard approach.
}


\begin{document}
\maketitle
\flushbottom

\section{Introduction \label{sec:introduction}}

Improvements in the precision high energy collider experiments are putting increasing pressure on theoretical
predictions. The latest technology for the evaluation of scattering amplitudes, the handling of infrared
singularities and Monte Carlo event generation has been able to achieve an impressive range of
predictions for differential observables at NLO and NNLO in both QCD and EW coupling expansions.
Despite the successes, simulations at the cutting edge of the precision frontier are often extremely
computationally expensive.

In this article we explore one way in which popular computer science technology can be used to
decrease the computational cost of precision simulations. In particular, we consider high
multiplicity scattering processes, with extremely high mathematical complexity, where it is less
clear how to make use of conventional interpolation methods such as polynomial fits and
interpolation grids \cite{Czakon:2008zk, Borowka:2016ehy, Heinrich:2017kxx, Jones:2018hbb,
Heinrich:2019bkc}. Neural networks, however, are by now a standard tool within the data analysis,
data science and machine learning communities and offer a general, non-linear parametrisation which
have the potential to approximate any continuous function~\cite{Cybenko1989}, and therefore could be useful in the
context of high multiplicity scattering.

\begin{figure}
\centering
    \includegraphics[width=0.7\textwidth]{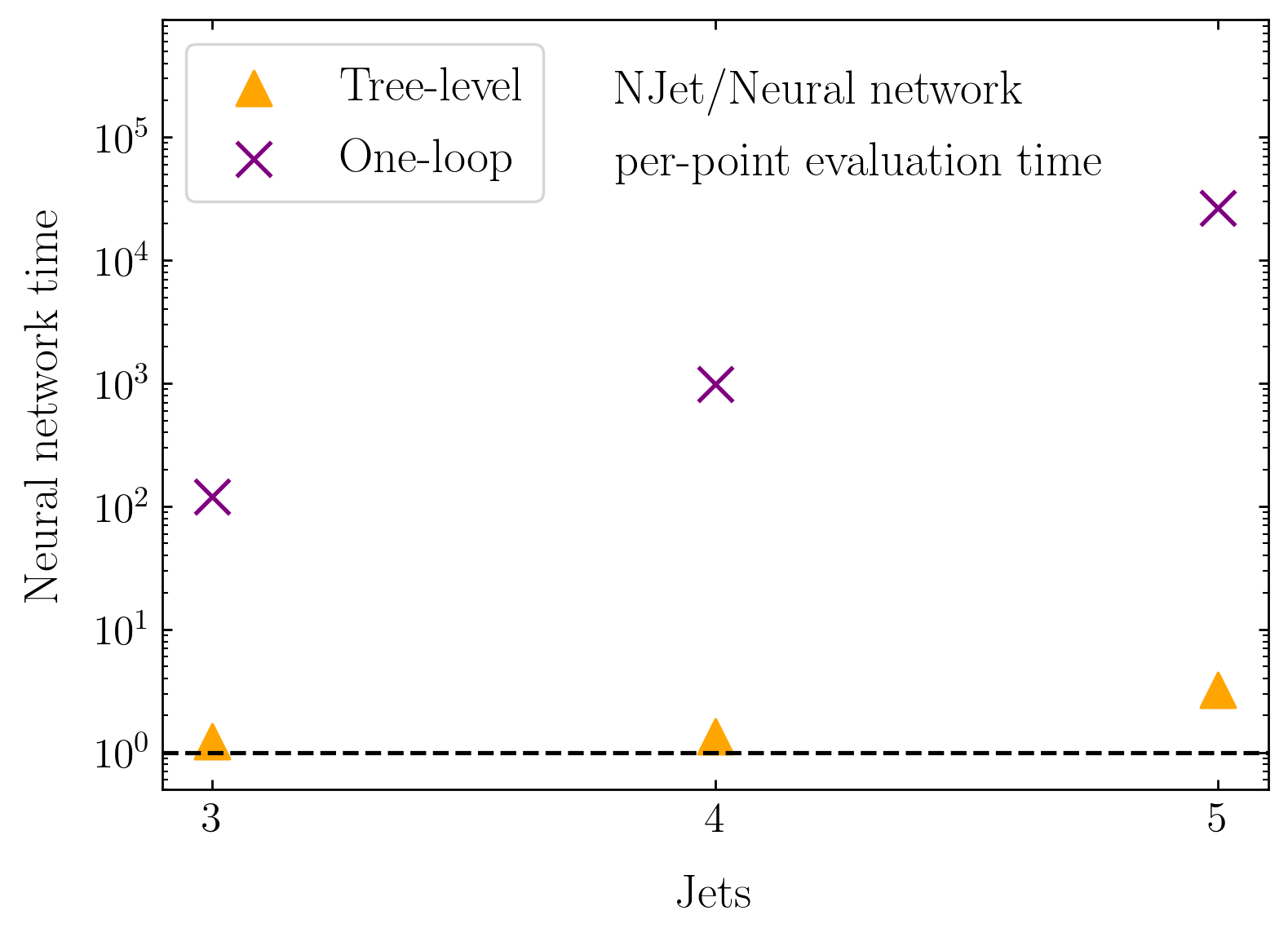}
    \caption{A comparison of the CPU cost of tree-level and one-loop amplitudes to a generic neural
    network (Keras/TensorFlow) as a function of the number of legs (equivalently number of
    variables). This demonstrates the very obvious fact that the neural network is fast to call
    and has a very mild dependence on the number of variables. The challenge is to train the network
    well enough that it can be interpolated and extrapolated reliably over a complete range of
    differential observables.}
\label{fig:timing}
\end{figure}

The basic principle is not new of course. Neural networks have the potential to provide extremely fast and lightweight
approximations of complicated amplitudes. In Figure \ref{fig:timing} we demonstrate this for the
particular test cases which are the subject of this article, the tree-level and one-loop amplitudes
inside the \njet~amplitude generator~\cite{Badger:2012pg} for $e^+e^- \to \leq 5$ jets. While the potential speed up in
the function call is quite striking, the real challenge is not clear from this analysis. The actual
improvement in CPU cost must take into account the time taken to train the network to a level that
it can be interpolated and extrapolated accurately and reliably.

Previous attempts have been made to use machine learning tools such as Boosted Decision Trees (BDTs) and neural networks for efficient phase-space sampling and Monte Carlo integration \cite{Bendavid:2017zhk, Klimek:2018mza} with recent work \cite{Bothmann:2020ywa, Gao:2020zvv} focusing on the use of coupling layers \cite{Dinh2014NICENI}. Similarly, work such as that of Otten \textit{et al.} \cite{Otten:2018kum} makes use of neural networks for explicit cross-section prediction. Here, the authors focus on $pp\to2$ jet processes, and implement an Artificial Neural Network point selection (NNPS) scheme for selecting training data based on the points the network struggles to learn the most. In addition, there has been much work on the use of Generative Adversarial Networks (GANs) \cite{NIPS2014_5423}, and other generative models, for event generation \cite{Otten:2019hhl, Hashemi:2019fkn, DiSipio:2019imz, Butter:2019cae, Butter:2019eyo, Carrazza:2019cnt, SHiP:2019gcl}, while there has been little work addressing the issue of explicit matrix element approximation \cite{Bishara:2019iwh}.


In this paper we design a deep learning pipeline to approximate $e^+e^-\to\leq5$ jet matrix elements
at both LO and NLO, thus exploring processes with significantly higher multiplicity than those
previously considered. While \cite{Otten:2018kum} uses a more automated approach for phase-space
sampling to aid in training a neural network, we employ physics-based knowledge of the processes in designing our pipeline and analyse
the effectiveness of this approach and what this might tell us about the phenomenological set up. We
pay careful attention to the errors and uncertainties in our neural network approximation, and offer
a comprehensive implementation of neural network regression analysis.

For usability, we supply code to accompany our methodology and results \cite{github}.

\section{Computational setup \label{sec:setup}}

We use the on-shell based C++ code \njet~\cite{Badger:2012pg}, to evaluate colour and helicity
summed born and virtual matrix elements for $e^+e^-\to \leq 5$ jets, denoted $\mathcal{M}^{(n,0)}$
and $\mathcal{M}^{(n,1)}$ respectively. \njet~uses integrand level reduction~\cite{Ossola:2006us}
and generalised unitarity \cite{Bern:1994cg, Britto:2004nc, Ellis:2007br, Giele:2008ve,
Forde:2007mi, Berger:2008sj, Badger:2008cm} to construct loop amplitudes from tree-level input
computed efficiently with Berends-Giele recursion~\cite{Berends:1987me}.  For a given phase-space
point, \njet~calculates the virtual and born matrix elements, along with the 1/$\epsilon$ and
1/$\epsilon^2$ correction coefficients, from which we can calculate the k-factors:

\begin{equation}
\text{k-factor} = \frac{\mathcal{M}^{(n,1)}}{\mathcal{M}^{(n,0)}}.
\end{equation}

For ease of use, \njet~is interfaced with via the Binoth Les Houches Accord (BLHA)
\cite{Binoth:2010xt}, which is designed to provide a standardised interface between Monte Carlo
tools and matrix element programs.

We explore the performance of various neural network parameterisations of the amplitude for total
and differential cross-section computations at LO, as well as their corresponding k-factor equivalents at NLO. We find that as the multiplicity increases, infrared
singularities on the edge of the phase-space increasingly cause problems for a single neural network, which
struggles to find a good fit across the whole phase-space. To improve the approximation, we
divide up the phase-space into sectors according to the subtraction method developed by Frixione, Kunszt and Signer (FKS) \cite{Frixione:1995ms,Frederix:2009yq}. Although we do not actually perform subtraction, this phase-space decomposition isolates the infrared singularities and allows the training of networks focused on improving their performance on each partition individually. 

\subsection{Phase-space partitioning \label{subsec:partition}}

We explore two pipeline configurations: i) we naively train a single network over all sampled points in phase-space; ii) we divide the phase-space into divergent and non-divergent regions in an attempt to partially isolate the infrared singularities and then further sub-divide the divergent region according to the FKS subtraction method, training one network on the non-divergent region, and a different network on each partition. For clarity we will generally refer to the single network, and ensemble of networks, as `models' and the individual networks comprising these models as `networks'.

We parameterise our phase-space according to the Lorentz invariant $y_{ij} = s_{ij}/s_{\text{com}}$, where $s_{ij} = (p_i + p_j)^2$ and $s_{\text{com}}$ is the centre-of-mass energy of the incoming particles, and define all cuts with respect to this quantity. Let the global phase-space cuts be denoted $y_{\text{cut}}$ and the partition dividing divergent and non-divergent regions be at $y_p$. Using these two scales, the divergent region, $\mathcal{R}_{\text{div}}$, and the non-divergent region, $\mathcal{R}_{\text{non-div}}$, are defined as follows:
 
\begin{equation}
\mathcal{R}_{\text{div}} = \{p\,|\, y_{\text{cut}} \leq \text{min}(y_{ij}) \leq  y_{\text{cut}} + y_p,\,p = (p_a,p_b,p_1,...,p_n),\,i,j \in \{1,...,n\}\},
 \end{equation}
 \begin{equation}
\mathcal{R}_{\text{non-div}} = \{p\,|\, y_{\text{cut}} + y_p \leq \text{min}(y_{ij}),\,p = (p_a,p_b,p_1,...,p_n),\,i,j \in \{1,...,n\}\},
 \end{equation}
 

 

where $p$ is a phase-space point consisting of the initial state 4-momenta, $p_a$ and $p_b$, and the outgoing momenta, $\{p_1,p_2,...,p_n\}$, where $n$ is the number of jets.

In the FKS subtraction formalism, the phase-space is divided such that the kinematic regions resulting from each partition contain only a specific subset of singularities. In order to achieve this, a set of ordered pairs, known as FKS pairs, are introduced. In our case of $e^+e^-\to \leq 5$ jets we define these as:

\begin{multline}\label{FKS pairs}
\mathcal{P}_{\text{FKS}} = \{(i,j)\,|\,3 \leq\ i \leq n_g+2,\,\,3 \leq\ j \leq n_g+2, i \neq j, \\
\mathcal{M}^{(n,0)}\,\, \text{or} \,\,\mathcal{M}^{(n,1)}\to\infty\,\, \text{if} \,\,p_i^0\to0\,\, \text{or} \,\,p_j^0\to0\,\, \text{or} \,\,\vec{p}_i||\vec{p}_j\},
\end{multline}

where $n_g$ is the number of gluons in the process. 

We then construct a partition function similar to that of \cite{Czakon:2014oma, Bartel:1986ua} (for a brief introduction to different FKS pair definitions and partition choices see Appendix \ref{AppendixB}):

\begin{equation}\label{S_i,j definition}
\mathcal{S}_{i,j} = \frac{1}{D_{1}s_{ij}}, \,\,\,\,\,D_1 = \sum_{i,j \in \mathcal{P}_{\text{FKS}}} \frac{1}{s_{ij}},
\end{equation}

such that:

\begin{equation}\label{FKS_cs_add}
\text{d}\sigma^{(X)} = \sum_{i,j} S_{i,j}\, \text{d}\sigma^{(X)},
\end{equation}

where, in this example, $\sigma^{(X)}$ represents either the Born cross-section, $\sigma^{(B)}$, the virtual correction, $\sigma^{(V)}$, or the k-factor, $\sigma^{(K)}$.

\begin{figure}
\centering
    \includegraphics[width=0.7\textwidth]{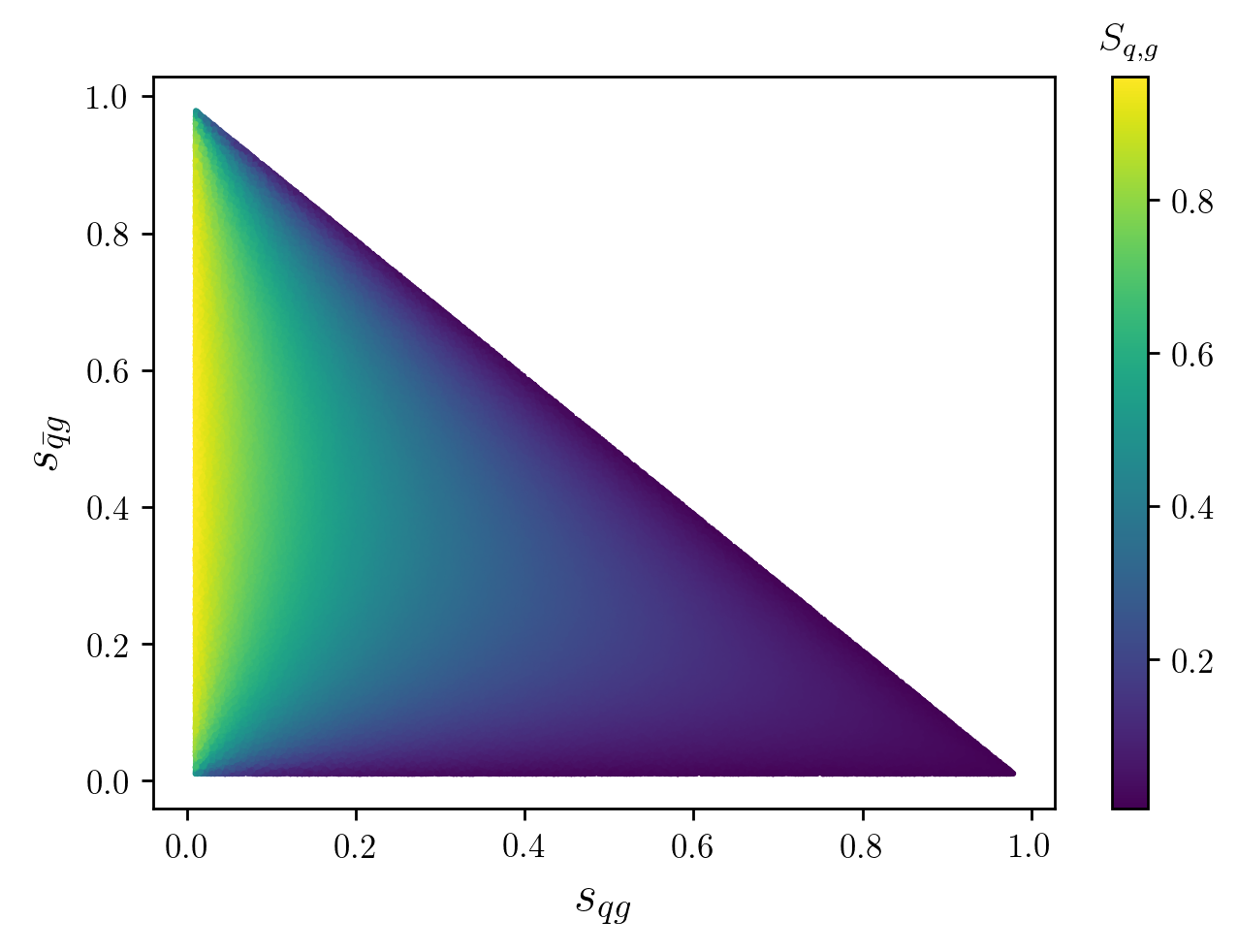}
    \caption {Behaviour of the $S_{q,g}$ FKS partition function}
\label{S_ij_qg}
\end{figure}

To demonstrate this partitioning effect we analyse the process $e^+e^-\to q\bar{q}g$. Here, we can isolate each of the two FKS pairs $\{qg, \bar{q}g\}$ and weight all phase-space points in the divergent regions according to the behaviour of $\mathcal{S}_{i,j}$ for each pair. The first pair corresponds to either the quark and gluon going collinear or the quark or gluon going soft. Since we cannot have soft quarks, this FKS partition only contains the singularities for the soft gluon and collinear quark and gluon. The behaviour of the FKS partition function, $S_{q,g}$ can be clearly seen in Figure \ref{S_ij_qg}, where we observe increasingly highly weighted points as $s_{qg}$ approaches 0. 

An advantage of this method is that the interpolation between singular regions is smooth since they add together to produce the overall cross-section (see Equation \ref{FKS_cs_add}).\footnote{An alternative implementation would be to partition the phase-space in a piecewise manner according to Heaviside step functions (as in \cite{Frixione:1995ms}); however, this introduces an additional set of scale choices and significantly reduces the number of phase-space points left for each network to learn the complicated divergent structure. Indeed, we found that when partitioning piecewise the network performs significantly worse in comparison to this smooth implementation.} By weighting the matrix elements in this way, phase-space points closer to the $q||g$ singularity contribute with increasing significance to the corresponding neural network's loss during training. A similar analysis can be performed for the second FKS pair in this process.

Since the FKS pairs are ordered, the upper bound on the number of pairs for our processes is:

\begin{equation}\label{N_max}
N_{\text{max}} = \frac{n_j(n_j-1)}{2} -1,
\end{equation}

where $n_j$ is the number of jets and the $-1$ comes from the fact that $\{q\bar{q}\}$ is not an FKS pair by definition. It should be noted that the number of pairs can be reduced in reality due to the symmetric behaviour of all gluon-gluon, or quark-gluon pairs; however, for simplicity we partition into $N_{\text{max}}$ regions. For example, in the example above, $N_{\text{max}} = 2$ but $N = 1$ since the behaviours of the two pairs in this process are identical.

After using the FKS partition function to divide the region $\mathcal{R}_{\text{div}}$, we are left with $N_{\text{max}} + 1$ regions in total across which we train the same number of networks. We find that setting the scale to $y_p = 0.01$ is generally applicable to all processes analysed.

\subsection{Neural network setup}

We compare the performance of two neural network setups, firstly a singular network is trained over the entire uniformly sampled phase-space, and secondly an ensemble of $N_{\text{max}}+1$ networks are trained over the partitioned phase-space.

\subsubsection{Data}

The phase-space is uniformly sampled using the RAMBO algorithm \cite{Kleiss:1985gy}, with each point initially having a weighting of unity. At LO, we train the single network model on data generated from sampling over the entire phase-space uniformly, whereas we train the ensemble model on samples drawn equally from the divergent and non-divergent regions. \footnote{Testing was done to assess the significance of equally sampling from the divergent and non-divergent regions of phase-space when training the single network approach as well, although we found little significant performance increase relative to that of using the ensemble approach.} At NLO, due to the computational expense of virtual matrix element calculation, the phase-space is uniformly sampled as a whole and then divided into $\mathcal{R}_{\text{div}}$ and $\mathcal{R}_{\text{non-div}}$ regions after sampling. RAMBO was chosen for its simplicity, the ease with which it can be altered to our specifications, and because it highlights interesting pitfalls and difficulties in high-dimensional functional approximations (see more on this below). In total we generate 500k phase-space points for training at LO, but only 100k at NLO due to the complexity of the problem.

The infrared poles in the matrix element result in singularities. Neural networks for classification tasks have been repeatedly shown to perform better when datasets are balanced, thus helping to avoid bias in the classification. Balancing can be done through a variety of methods such as over and under sampling, as well as loss function weightings. In regression tasks, the equivalent to class imbalances are under sampled regions that behave significantly differently to the rest of the sampled space. When doing explicit numerical calculations of the matrix element, these imbalances are not such an issue and their effect when calculating observables can be estimated by the Monte Carlo error and by phase-space resampling, yet they become significant when training a network. Through balancing the training datasets in the divergent and non-divergent regions, and using the FKS partitioning method as outlined above, we hope to address the issue of underrepresented regions.

Increasingly sophisticated non-machine learning based methods for phase-space sampling using adaptive methods \cite{Friedman, Lepage:1977sw, Lepage:1980dq, Ohl:1998jn, Kroeninger:2014bwa}, including the use of recursive stratified sampling \cite{Press:1989vk} and integrand factorisation \cite{Jadach:2002kn} have been developed. Similarly, importance sampling methodologies specifically designed for QCD antenna generation exist to better capture these divergent regions given the physical knowledge of the pole structure \cite{Draggiotis:2000gm, vanHameren:2002tc}. RAMBO, however, is indifferent to these variational differences in phase-space, giving a more naive sampling, yet the ability to construct an interpolation function from a uniformly sampled phase-space means we save computational time during the sampling stage. Although performance of our approximation may be increased using these more sophisticated methods, demonstrating sufficiently good results while requiring only the use of simple sampling techniques like RAMBO further shows the power of our method and the additional time savings it can offer.

Once the phase-space points are generated, we use \njet~\cite{Badger:2012pg} to calculate the corresponding squared matrix elements at LO, and the virtual correction terms at NLO, for $e^+e^-\to Z^*/\gamma\to q\bar{q} + n_g$. We calculate all quantities in the four-dimensional helicity (FDH) scheme, assuming all external legs to be massless, with the number of light quark flavours set to $n_f = 5$, and we use the same renormalisation scale as in \cite{Frederix:2010ne} (see more details in Section \ref{sec:errors_analysis}). 

When training the network, the dataset is split in an 80:20 ratio for training and validation. Furthermore, independently generated, unseen datasets are used in testing. Through generating many more points for testing than training we demonstrate both the performance of our methodology as an interpolation function, as well as its ability to capture the increasingly sampled divergent structure.

To avoid the problem of vanishing/exploding gradients, we standardise our data to zero mean and unit variance at each input node and across the targets.

\subsubsection{Architecture}

Choosing an optimal network architecture is non-trivial due to the large number of parameters that can be tuned to an array of criteria. It is common to approach a singular problem using a neural network and thus optimise the architecture for that process; however, while we want to demonstrate the ability of networks to become sophisticated multi-parameter interpolation functions, we require these models to generalise to a variety of processes.

For better generalisability we do not fine-tune a network to any particular process, but rather attempt to employ the same architecture for each process. The neural networks are parameterised using Keras \cite{keras} with a Tensorflow \cite{tensorflow} backend and comprise of fully-connected layers with an input layer of $(n_j-1)\times 4$ nodes and output of 1, with three hidden layers made up of of 20-40-20 nodes. The hidden layers all use hyperbolic tangent activation functions and the output node has a linear activation function.

The loss function is taken to be the mean squared error, 

\begin{equation}
L = \frac{1}{n}\sum^{n}_{i=1}(f(x_{i}) - y_i)^2,
\end{equation}

where $n$ is the number of training points, $f: \mathbb{R}^d\to\mathbb{R}$ is the function describing the neural network, $x_i$ is the ith $d$-dimensional input data, and $y_i$ the corresponding target variable. The network is optimised using Adam optimisation \cite{adam}, while the number of training epochs is determined through Early Stopping (see Section 8.1.2 \cite{goodfellow}), tracking the validation loss with no minimum change requirements. We recognise that by using a validation set containing only 20\% of the original training set, we may be severely limiting the number of points in the increasingly divergent regions, thus skewing our Early Stopping criteria to the less divergent regions. In an attempt to mitigate this, we train with a patience of 100 epochs to measure effects in the loss function significantly later in the training regime; however, at NLO we found that this makes minimal difference to the total loss and so can be reduced to speed up network training.

The inputs to the network are the 4-momenta of $n_j-1$ jets. Since we fix the centre-of-mass energy for training, we sought to reduce the number of input nodes for more efficient learning. We note that further reductions in the number of input parameters could be made, yet in testing this had no significant effect on performance.

\subsection{Uncertainty Analysis \label{sec:errors_analysis}}

The subject of error and uncertainty analysis in machine learning processes is receiving increasing attention (see \cite{tagasovska:2018, Gal:2016} and the references therein), especially in the particle physics community \cite{Nachman:2019dol, Nachman:2019yfl, Bollweg:2019skg, Englert:2018cfo}, yet too frequently a demonstration of rigorous error analysis in machine learning regression processes is lacking.

As stated in \cite{tagasovska:2018}, the main sources of error arise from approximation, aleatoric and epistemic uncertainties.\footnote{Approximation uncertainty arises due to the model being too simplistic to allow for complex functional fitting, e.g. too few nodes or hidden layers in a neural network meaning the model isn't able to fit sufficiently non-linear functions. Aleatoric uncertainty accounts for fluctuations in the data distribution e.g. from measurement errors, and cannot be decreased by collecting more data from the same experimental setup. Epistemic uncertainties, on the other hand, account for uncertainties in the model, including lack of sufficient coverage of the data.} Since we are using deep neural networks, and have tested both deeper and shallower architectural designs, we assume that errors associated with approximation uncertainties are negligible. Additionally, we do not consider aleatoric uncertainties here since our data has been generated through high-precision numerical methods. Moreover, \njet~accuracy tests have been performed to measure the stochasticity in matrix element generation and found this fluctuation to be negligible.\footnote{\njet~accuracy tests are performed by inferring on each phase-space point twice and checking the difference in the results. The threshold is set to the default value of $10^{-5}$ and errors arise due to lack of floating point precision and rounding errors.} Following \cite{Nachman:2019dol} we apply similar methods highlighted for use in classification networks, to this regression task. Specifically, we focus on the measurement of precision/optimality errors which include those arising due to epistemic uncertainties. 

We measure model parameter initialisation dependence by retraining models on fixed training datasets while randomly reinitialising the weights. Depending on the observable, the standard deviation in the bins can be measured. Additionally, when sampling the phase-space, the Monte Carlo error is calculated; however, this does not fully account for the uncertainty in phase-space completeness. For this we bootstap the training data thereby resampling the phase-space multiple times and retrain the networks on different datasets, while keeping the weight initialisations fixed. Since in this paper we are comparing neural network output against \njet~results, to avoid the double counting of errors we only include Monte Carlo errors on the \njet~results.

The performance of our methodologies are also dependent on the test set chosen. For this we quote the Monte Carlo error, although it should be noted that the same issue with determining sampling completeness occurs here. Due to the computational expense of repeated generation of test sets we do not perform this, although the uncertainty bands on the neural network approximations should be sufficient to provide evidence of our methodology since these additional dataset dependancies are negligible given the large number of test points used and the relative size of the computed Monte Carlo error compared to the network uncertainties (see Section \ref{sec:results}).

The errors on the network approximation that we calculate are therefore the error due to model initialisation dependence and error due to the size of the training dataset, which are added in quadrature. As noted by Nachman \cite{Nachman:2019dol}, additional sources of uncertainty are inherent in the network approximation that are hard to calculate explicitly, such as dependence on the model architecture (e.g. the number of hidden layers, nodes in each layer and the types activation functions used). Due to the size of the other errors mentioned, and the lack of currently available tools for their calculation, we do not attempt to incorporate errors arising from these uncertainties into our analysis. We quote Monte Carlo error only for the testing dataset.

When presenting our results, we calculate the mean of the multiple models trained and quote both the standard error on the mean, to demonstrate how uncertainty bands might be obtained in practice through the training of an ensemble of models where all models are used, and the standard deviation from the mean, showing the spread of the different trained models and the potential variability in performance if only a single model were to be trained. In an attempt to avoid confusion, we do not plot these uncertainties simultaneously, rather we are specific in figure descriptions as to which one we are plotting. Throughout this paper, we choose to train 20 models for each example, however, this number was chosen in a slightly \textit{ad hoc} manner, since it gave a reasonable distribution of models, and should not be interpreted as a requirement.

Theoretical uncertainties are also prevalent in all of these calculations due to variability in setting the renormalisation scale, $\mu$. Such uncertainties propagate though the networks since a model will learn to fit data at a certain scale. In this paper we train on data generated at a fixed scale as used in \cite{Frederix:2010ne}. We perform the normal \textit{ad hoc} scale variation of $\mu/2$ and $2\mu$ purely to determine the dependence of our methodology on such a scale choice. In doing so we found that the network is able to approximate the matrix elements at each scale equally well to within Monte Carlo error, and we therefore assert that the network performance is not highly dependent on the value of $\mu$ in the range we analysed. Moreover, since the goal of this work is not to calculate the cross-section or k-factors of a new process, but to provide tools for estimating such values for already known process, we do not quote these as uncertainties in our methodology.

\section{Results \label{sec:results}}

We test our methodology on estimating both LO cross-sections and k-factors for processes up to $e^+e^-\to5$-jets. In addition, various observables are plotted to demonstrate the applicability of our methodology to real calculations. In general, we see that neural network approximations demonstrate wide applicability to the cases investigated, with the FKS partitioning method giving more accurate and stable results due better approximating the infrared singularities.

It should be noted that to retain consistency between the training and testing phases, we sample both datasets in the same way, i.e. the test set used for single network inference has been sampled uniformly over the entire phase-space, whereas the set used in the ensemble case has been split into divergent and non-divergent regions and uniformly sampled equally and independently in each (as described in Section \ref{subsec:partition}). This means that, while the ensemble approach has received a greater number of divergent points during training at LO, by sampling the test set in the same way we hope to make the tests equally `difficult' in proportion to the sampling method used. Throughout all tests, phase-space cuts at $y_{\text{cut}}$ are used to regulate the infrared divergences.
\newpage

\subsection{Approximations at LO}\label{LO results}

Although leading order calculations are not significantly computationally expensive, they pose interesting test cases for neural network approximations of high multiplicity processes with many scales and complex infrared singularity structures. Moreover, we find that much of wha can be learnt from the performance of the models here can be applied to the NLO case.

As detailed above, we compare a single network trained over the entire phase-space with an ensemble of networks each trained on $N_{\text{max}} + 1$ partitions of phase-space. In determining the appropriate value of the global phase-space cut parameters, $y_{\text{cut}}$, we evaluate the performance of our models by calculating the ratio of the network output to the \njet~calculation as well as the network's ability to approximate the cross-section.


\begin{figure}
\centering
\begin{subfigure}{.49\textwidth}
    \includegraphics[width=\textwidth]{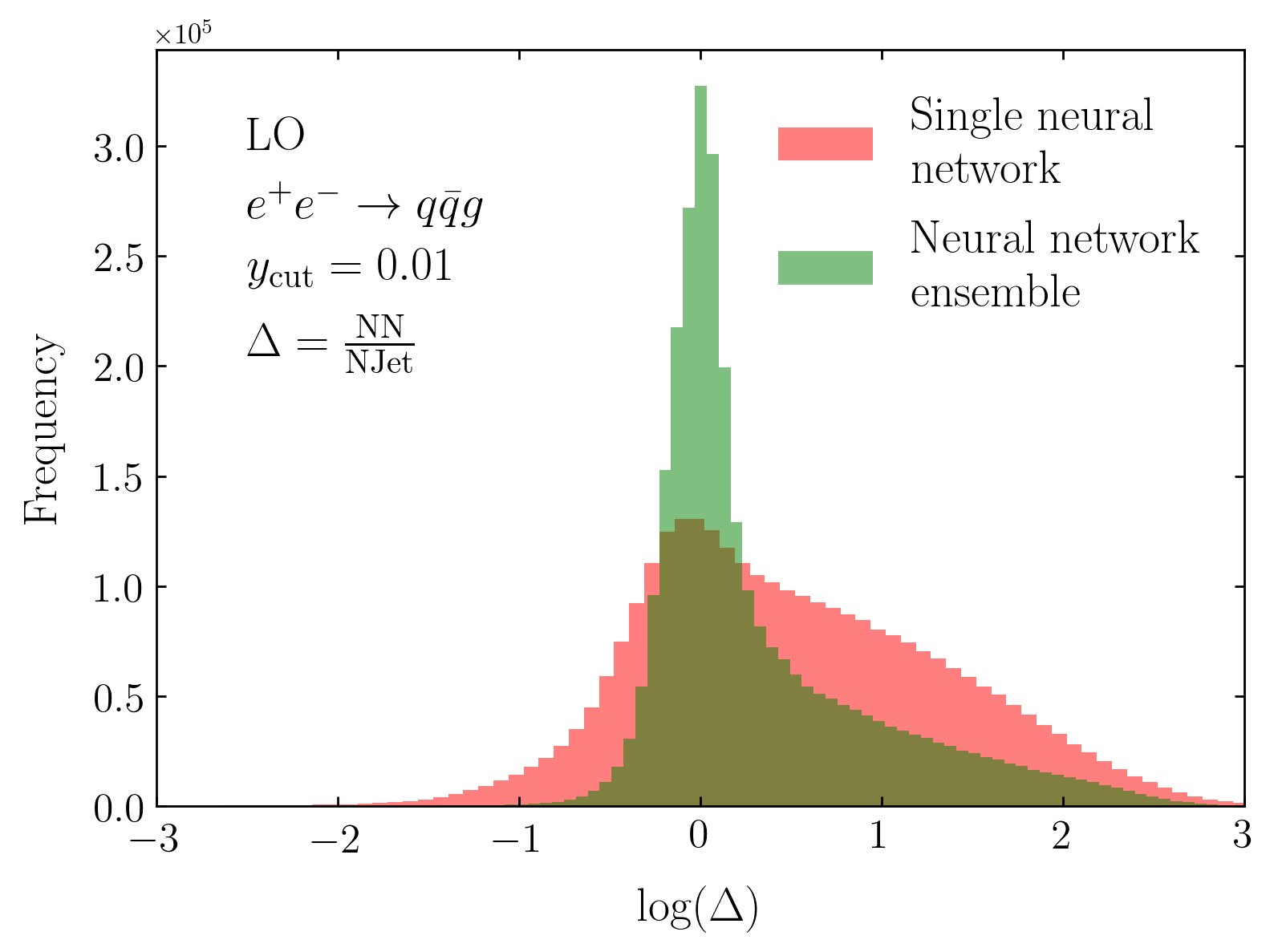}
\end{subfigure}%
\hfill
\begin{subfigure}{.49\textwidth}
    \centering
    \includegraphics[width=\textwidth]{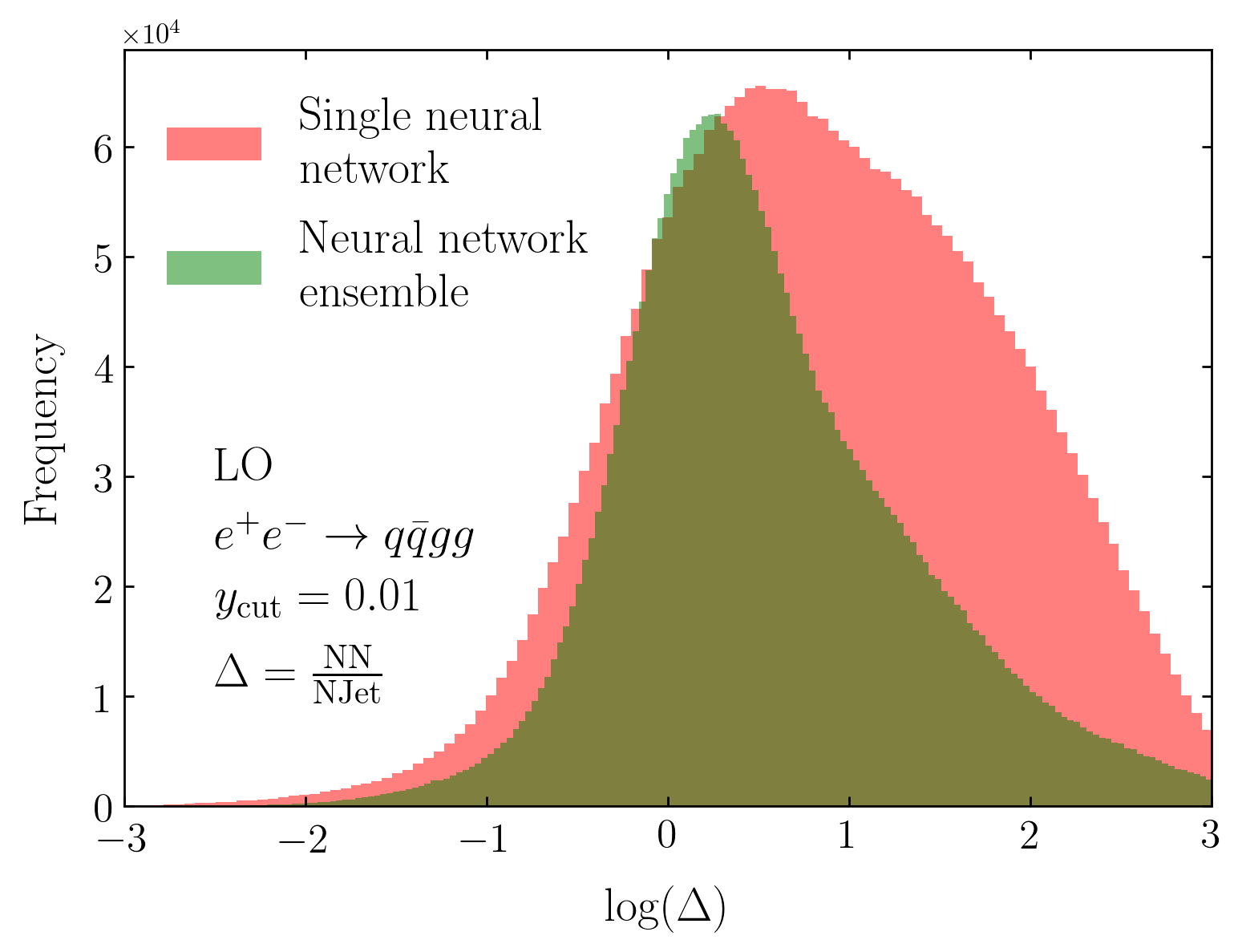}
\end{subfigure}

\centering
\begin{subfigure}{.49\textwidth}
    \includegraphics[width=\textwidth]{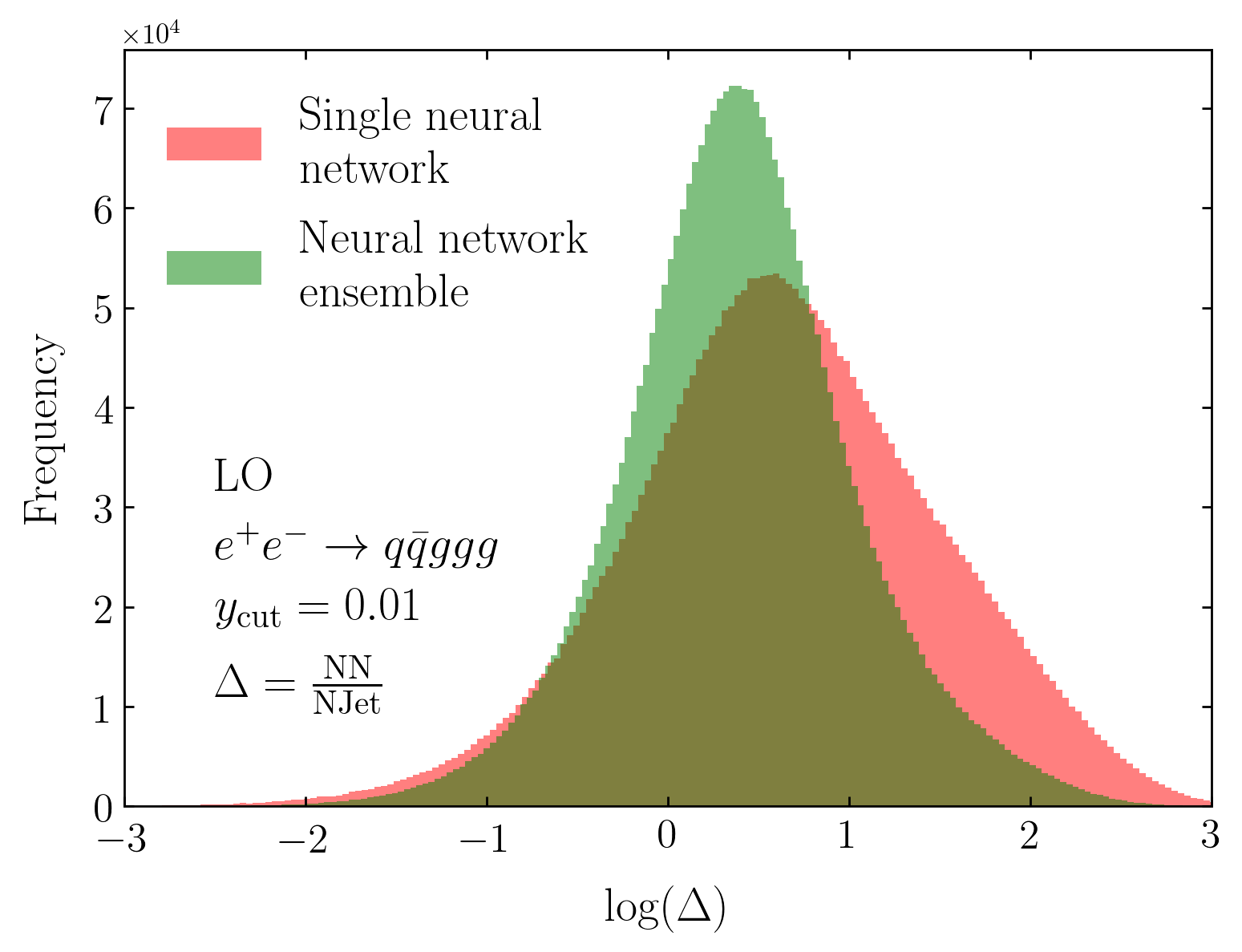}
\end{subfigure}%
\hfill
\begin{subfigure}{.49\textwidth}
    \centering
    \includegraphics[width=\textwidth]{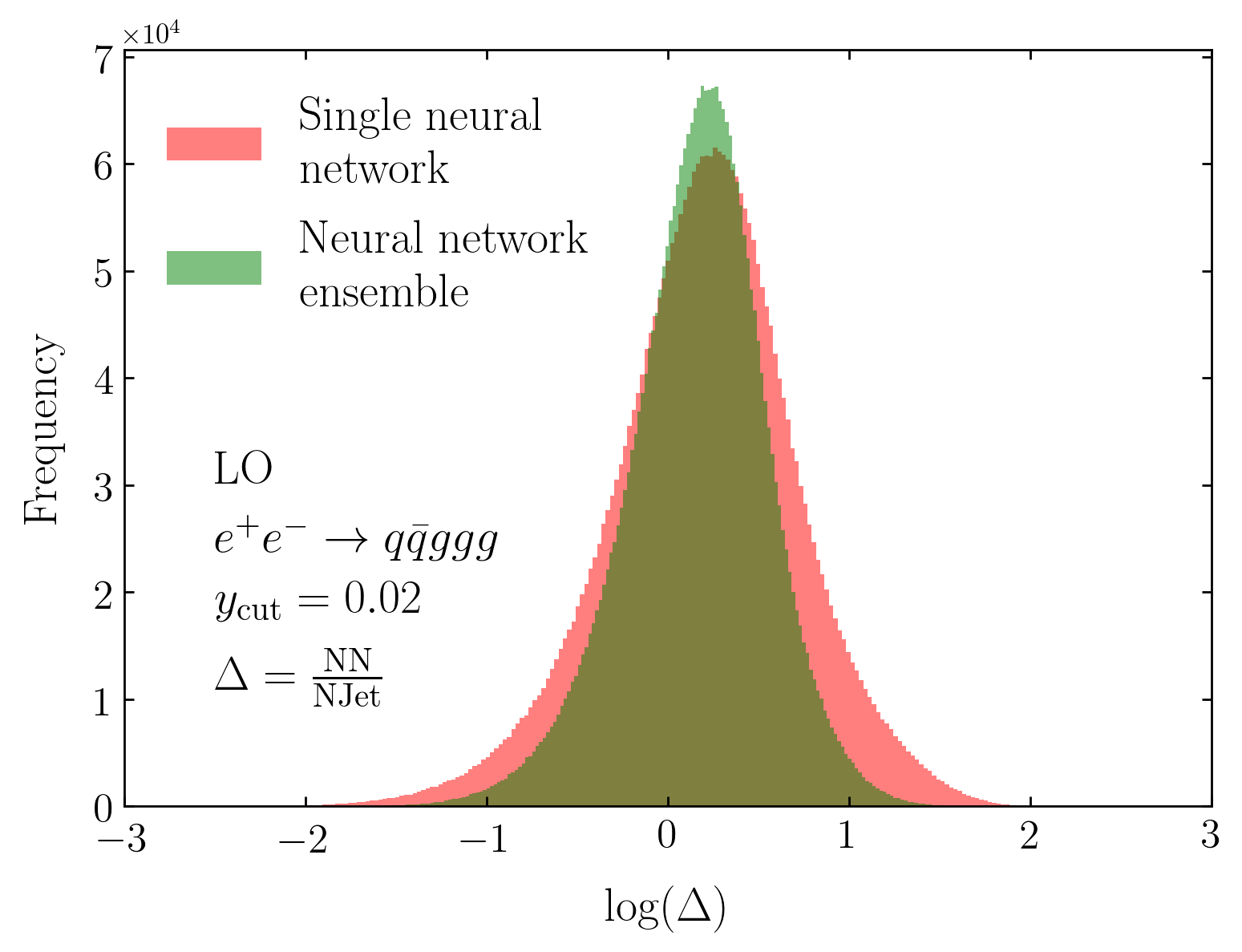}
\end{subfigure}

\caption{Born matrix element output of a single neural network (red) and our ensemble approach (green) compared to the \njet~calculation. Outputs are taken as the average over 20 trained models or ensembles.}
\label{LO_errors}
\end{figure}

Figure \ref{LO_errors} shows the distribution of the neural network errors by calculating the ratio of the model output and the \njet~result at each phase-space point in the test set. Since the ensemble of networks gives much narrower and more Gaussian shaped distributions than the single network approach, we can clearly see that this method is preferable at the level of per-point accuracy. Additionally, the error distributions of the ensemble approach are also more closely centred on zero, in comparison to the single network approximation, thus suggesting that the ensemble will also produce a better overall average performance as well. Note that these plots to not contain any information about the relative uncertainties attached to these model outputs, which we will discuss below.

\begin{figure}[H]
\centering
\begin{subfigure}{.49\textwidth}
    \includegraphics[width=\textwidth]{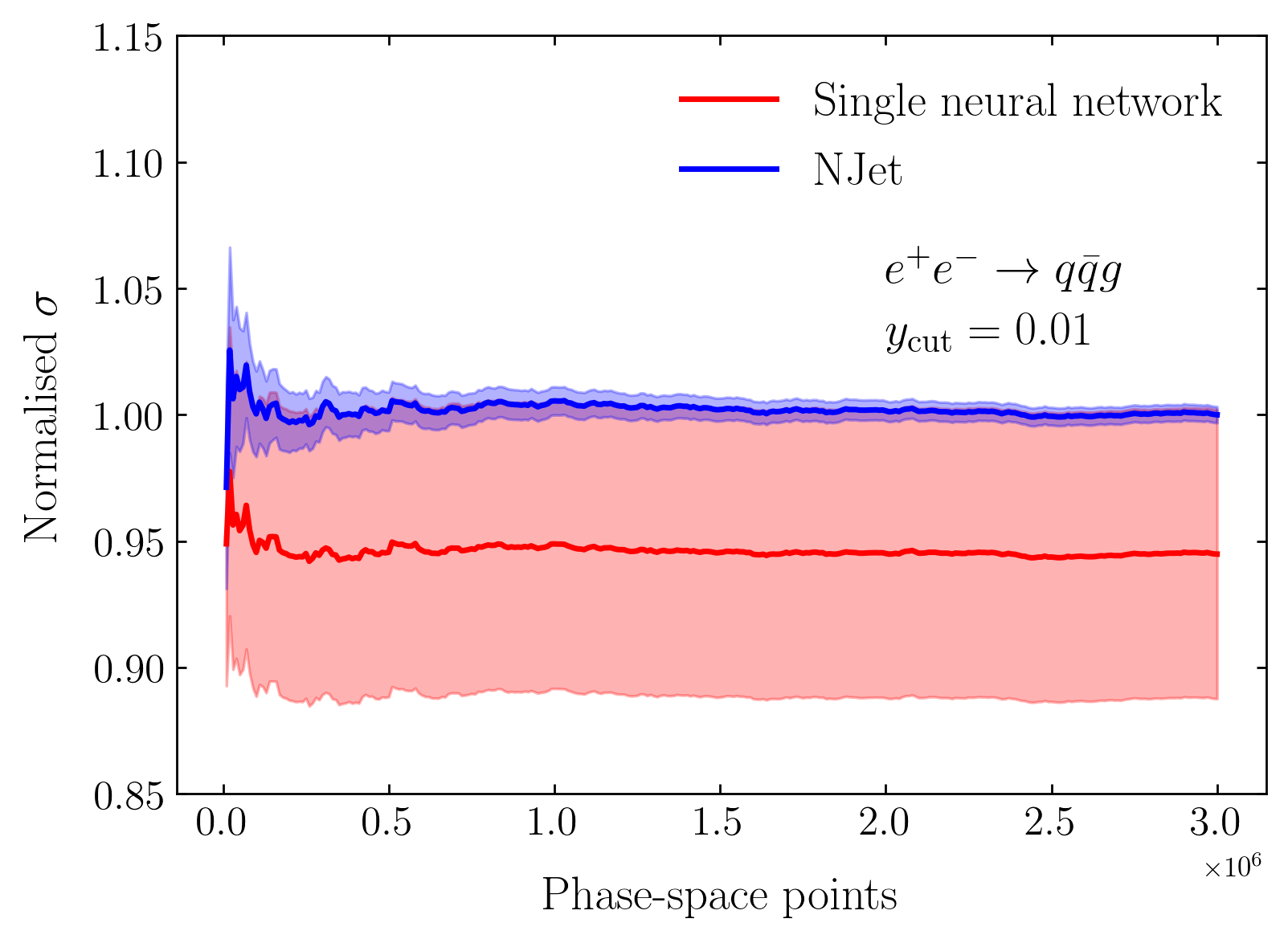}
\end{subfigure}%
\hfill
\begin{subfigure}{.49\textwidth}
    \centering
    \includegraphics[width=\textwidth]{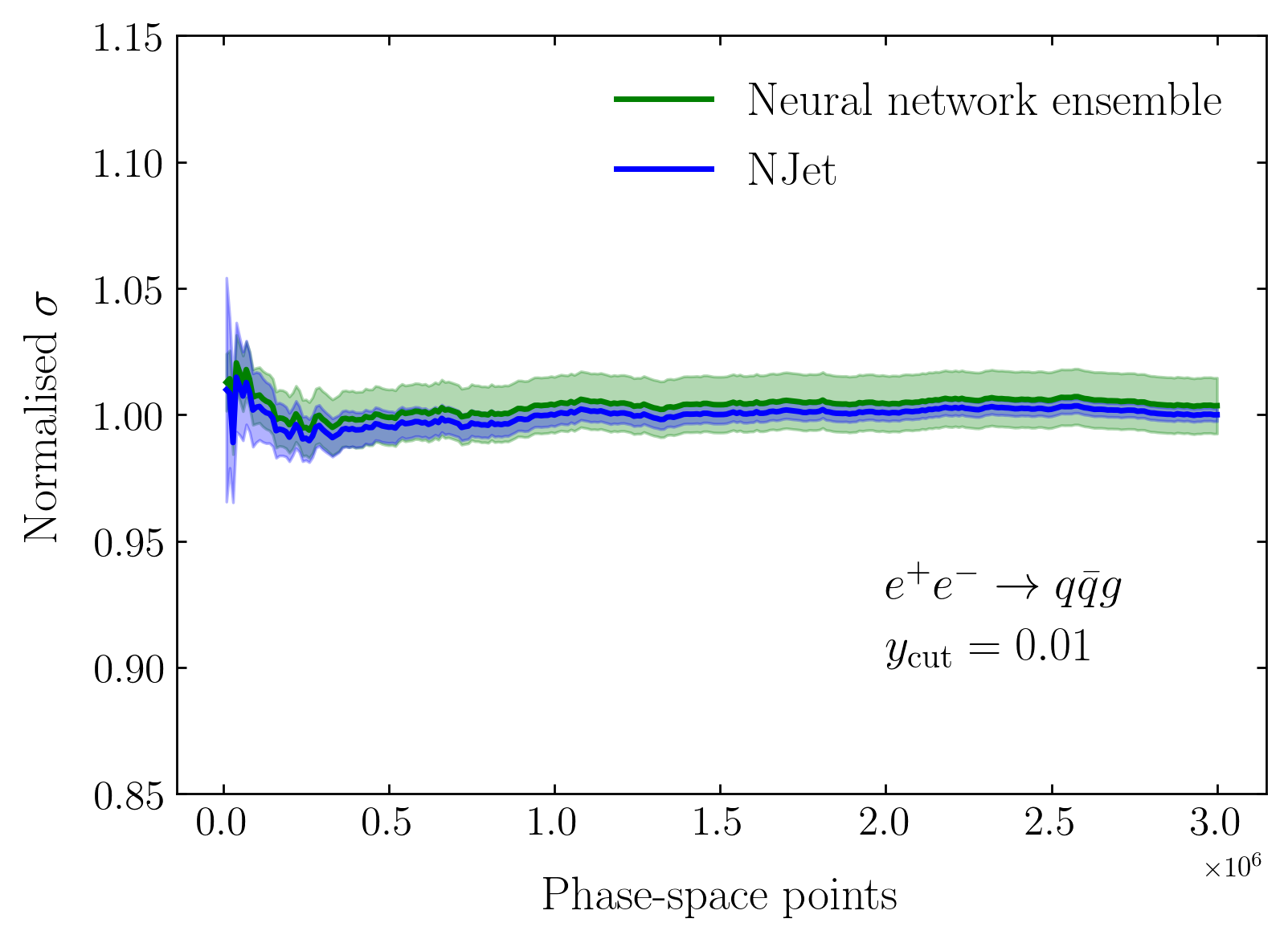}
\end{subfigure}

\centering
\begin{subfigure}{.49\textwidth}
    \centering
    \includegraphics[width=\textwidth]{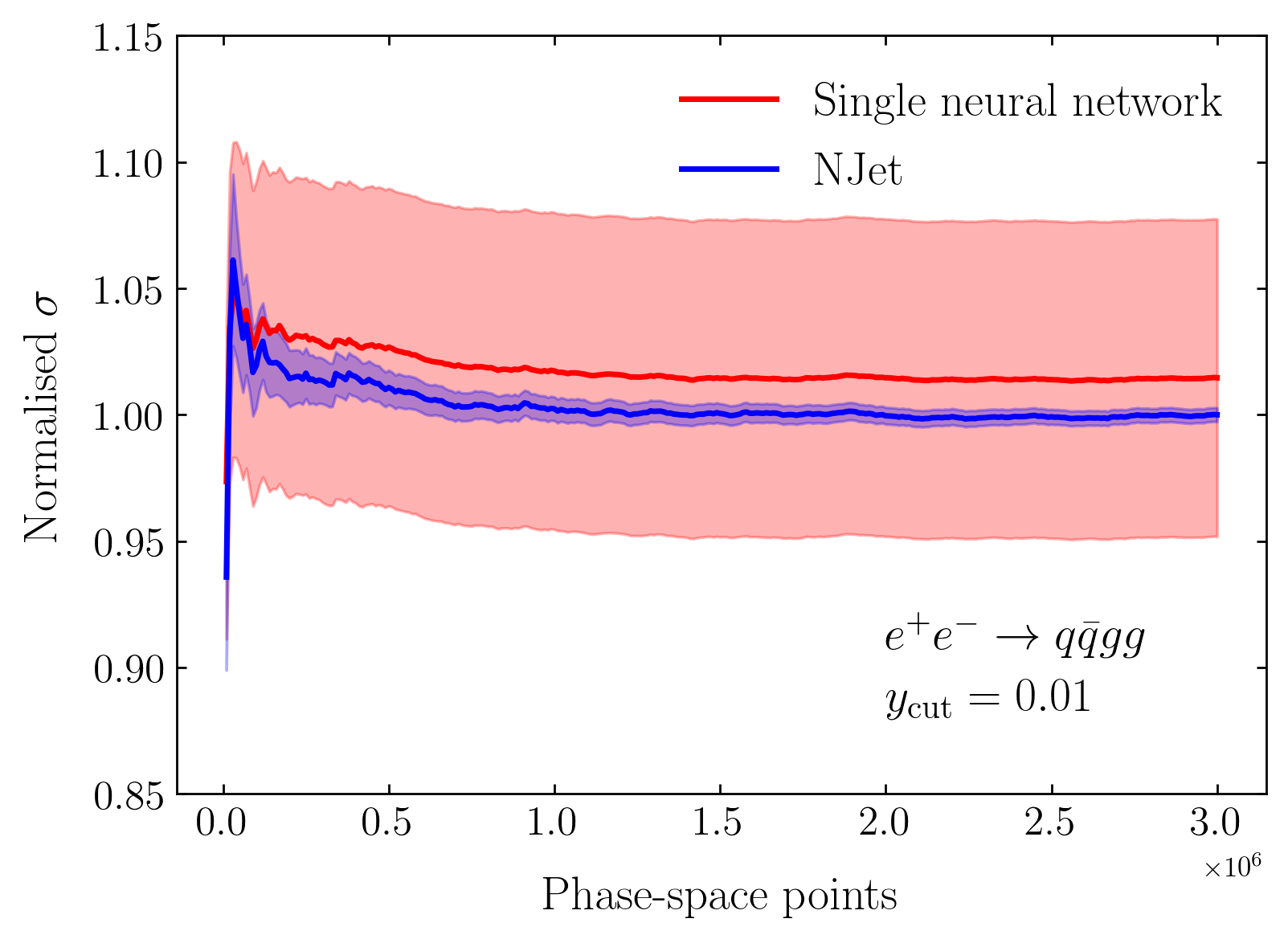}
\end{subfigure}%
\hfill
\begin{subfigure}{.49\textwidth}
    \centering
    \includegraphics[width=\textwidth]{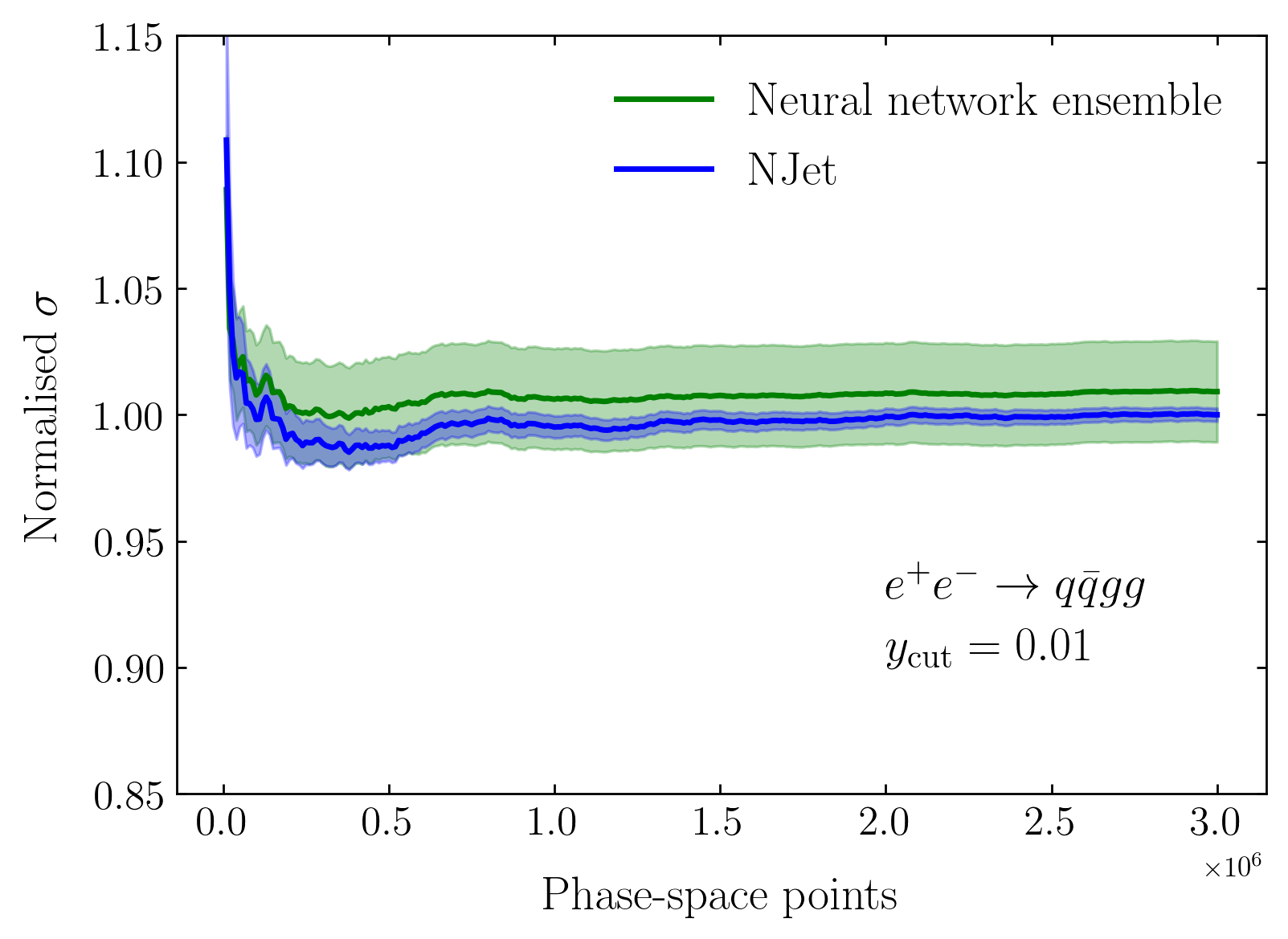}
\end{subfigure}

\centering
\begin{subfigure}{.49\textwidth}
    \centering
    \includegraphics[width=\textwidth]{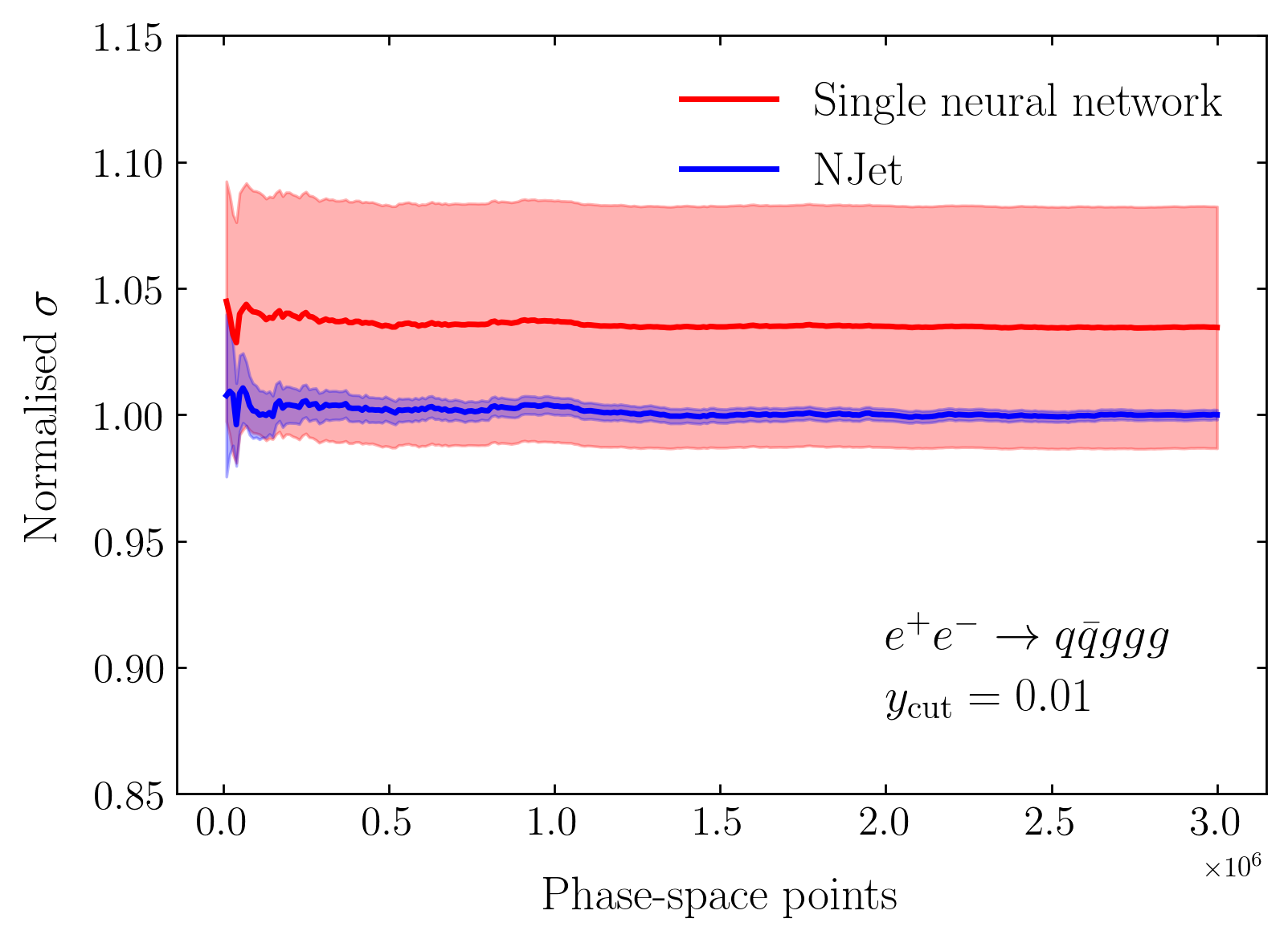}
\end{subfigure}%
\hfill
\begin{subfigure}{.49\textwidth}
    \centering
    \includegraphics[width=\textwidth]{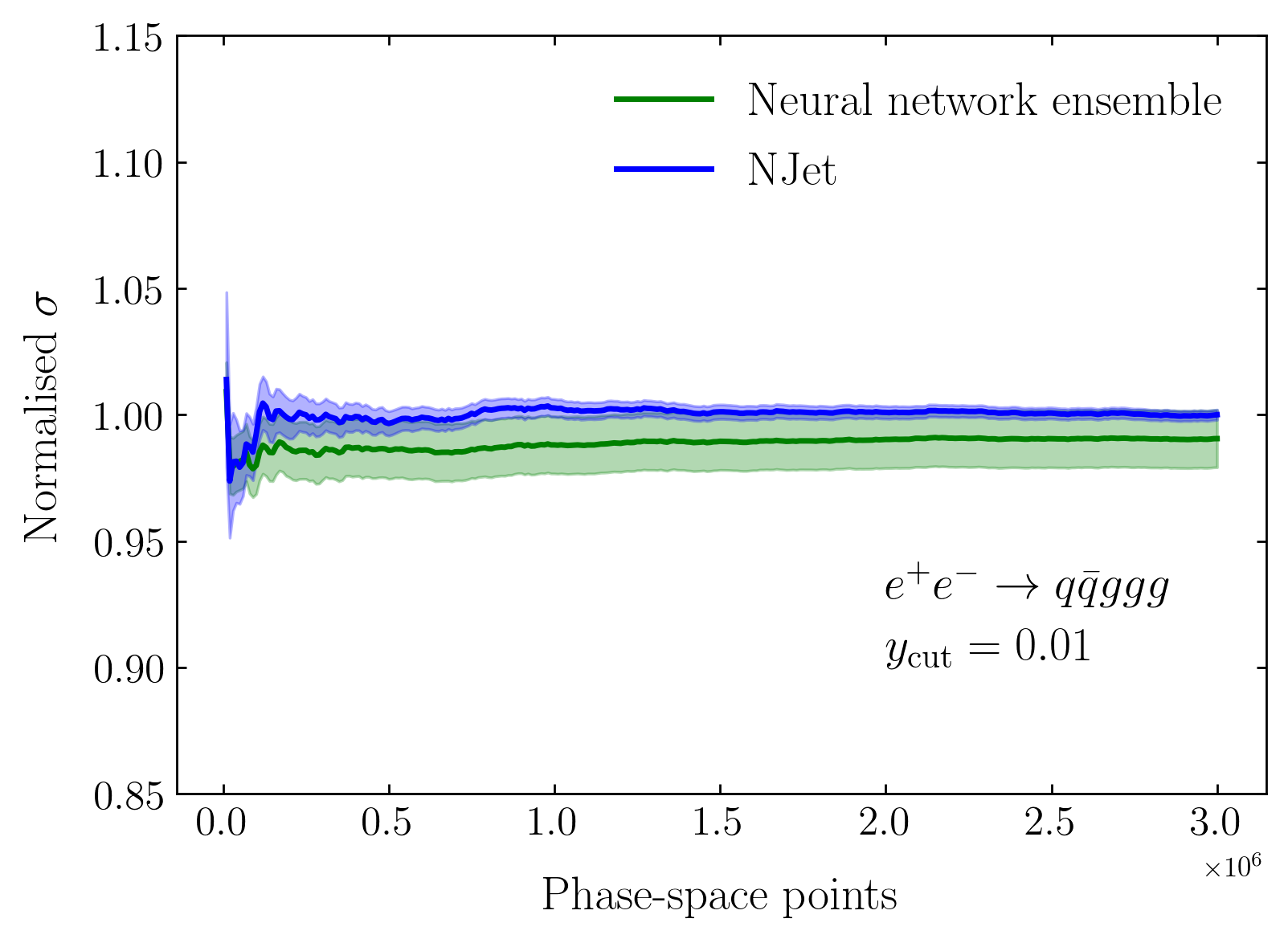}
\end{subfigure}

\centering
\begin{subfigure}{.49\textwidth}
    \centering
    \includegraphics[width=\textwidth]{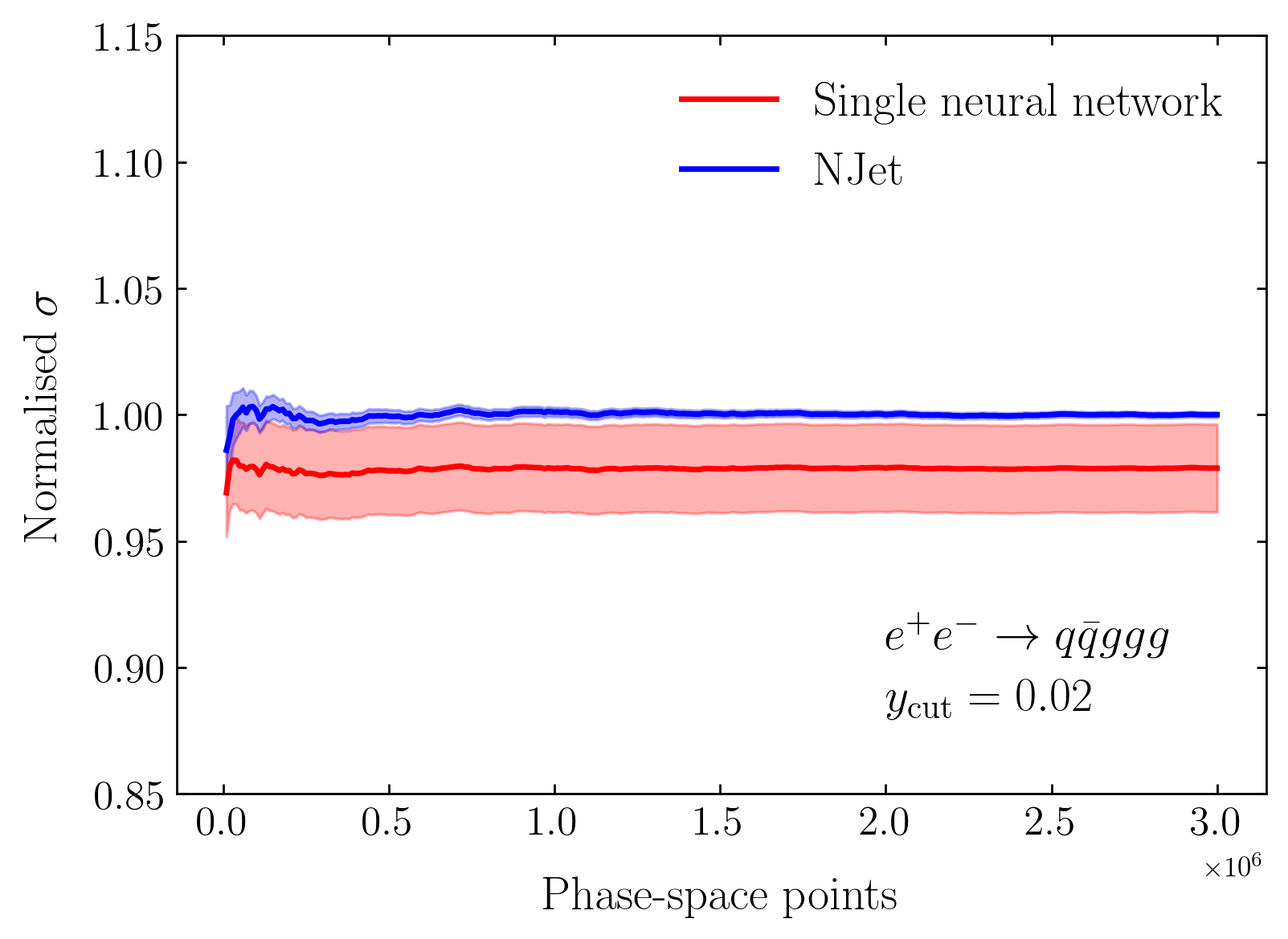}
\end{subfigure}%
\hfill
\begin{subfigure}{.49\textwidth}
    \centering
    \includegraphics[width=\textwidth]{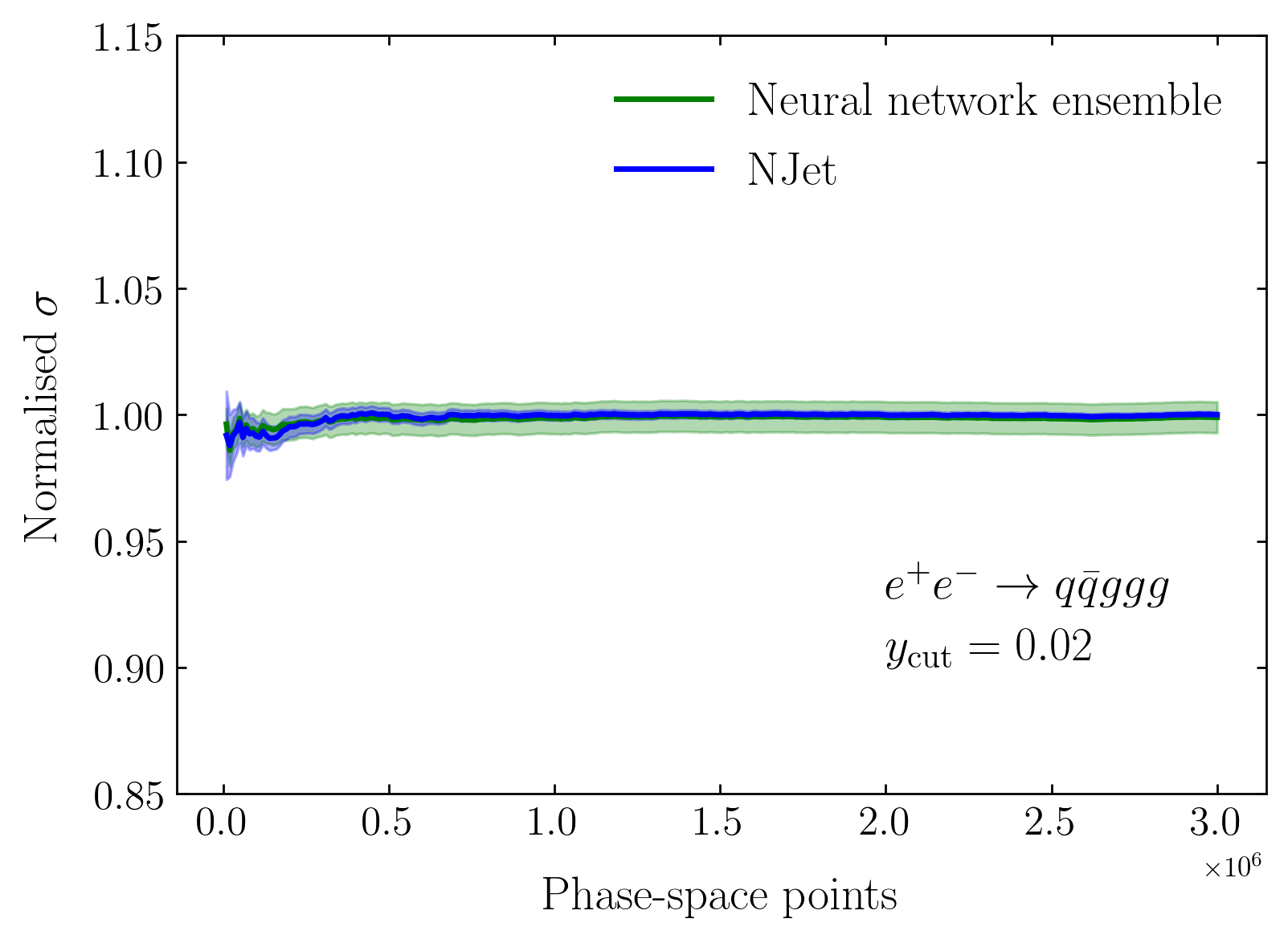}
\end{subfigure}
\caption{Comparison of a single neural network (left) vs. our ensemble approach (right) in estimating the Born normalised cross-section. Uncertainty bands denote the standard error on the mean calculated over 20 trained models (red and green) and Monte Carlo error on the \njet~result (blue).}
\label{LO_cs}
\end{figure}

While the error plots demonstrate the per-point performance of the models, we also wish to compare their performance in calculating physics observables while also taking into account uncertainty in the data and the model setup. Figure \ref{LO_cs} shows the approximated cross-sections of the naive and ensemble networks as compared to those computed from the \njet~matrix elements. As expected, we see a harsher $y_{\text{cut}}$ value at 5-jets better regulates the divergent regions, thus improving both the single and ensemble network approaches; however, this harsher cut is not fully necessary in the ensemble case as the \njet~result sits on the edge of the neural network uncertainty bands. 

When approximating the cross-section, we find the uncertainty bands to have very little noise and follow the shape of the average result closely. As we use a mean squared loss function for training, it can be shown that the network will tend towards learning the \textit{average} of the target distribution (see Appendix \ref{AppendixA}). Since no model will perfectly learn this distribution, for each model there will be an offset, $\epsilon$, between the final trained network average and the true distribution average. As the cross-section is proportional to the average over the phase-space, for any value of $n$, this offset will manifest itself as a distance away from the true cross-section such that:

\begin{align}
\frac{1}{n}\sum_{i = 1}^{n}(f(x_i) - y_i) &= \sigma_P - \sigma_N \label{XS no label} \\
& = \epsilon + \mathcal{O}(\theta)  \label{XS epsilon},
\end{align}

where $\sigma_P$ and $\sigma_N$ are the predicted and \njet~calculated cross-section values respectively, and $\theta$ is a small noise parameter. This therefore explains the relatively fixed distance between the model unceratinty upper and lower bounds and the \njet~result.

Another result of Equation (\ref{XS epsilon}) is that, unlike Monte Carlo error, inferring on more test points will not reduce the model uncertainties, since these uncertainties are intrinsically tied to the training set and the model initialisation. Any efforts to reduce these errors should therefore be focussed on addressing such uncertainties, as we do by developing our ensemble method, rather than focussing on the test dataset.

In general, the global cuts required for the ensemble approach to be within the Monte Carlo error of the true cross-section are $\sim\,y_{\text{cut}}=0.01$. These cut values are reasonable for our definition of $y_{ij}$ and are equivalent to the cuts made in \cite{Klimek:2018mza}.

After cuts have been made, we see that the ensemble of networks has a significantly reduced standard error when compared with the naive single network approach, with a predicted mean closer to the final stable cross-section. This difference in uncertainty can be understood by comparing the relative standard deviations of the single network and the deviations in the different networks making up the ensembles, as we shall now show.


\begin{figure}[H]
\vspace{-2.3cm}
\centering
\begin{subfigure}{.49\textwidth}
    \includegraphics[width=0.92\textwidth]{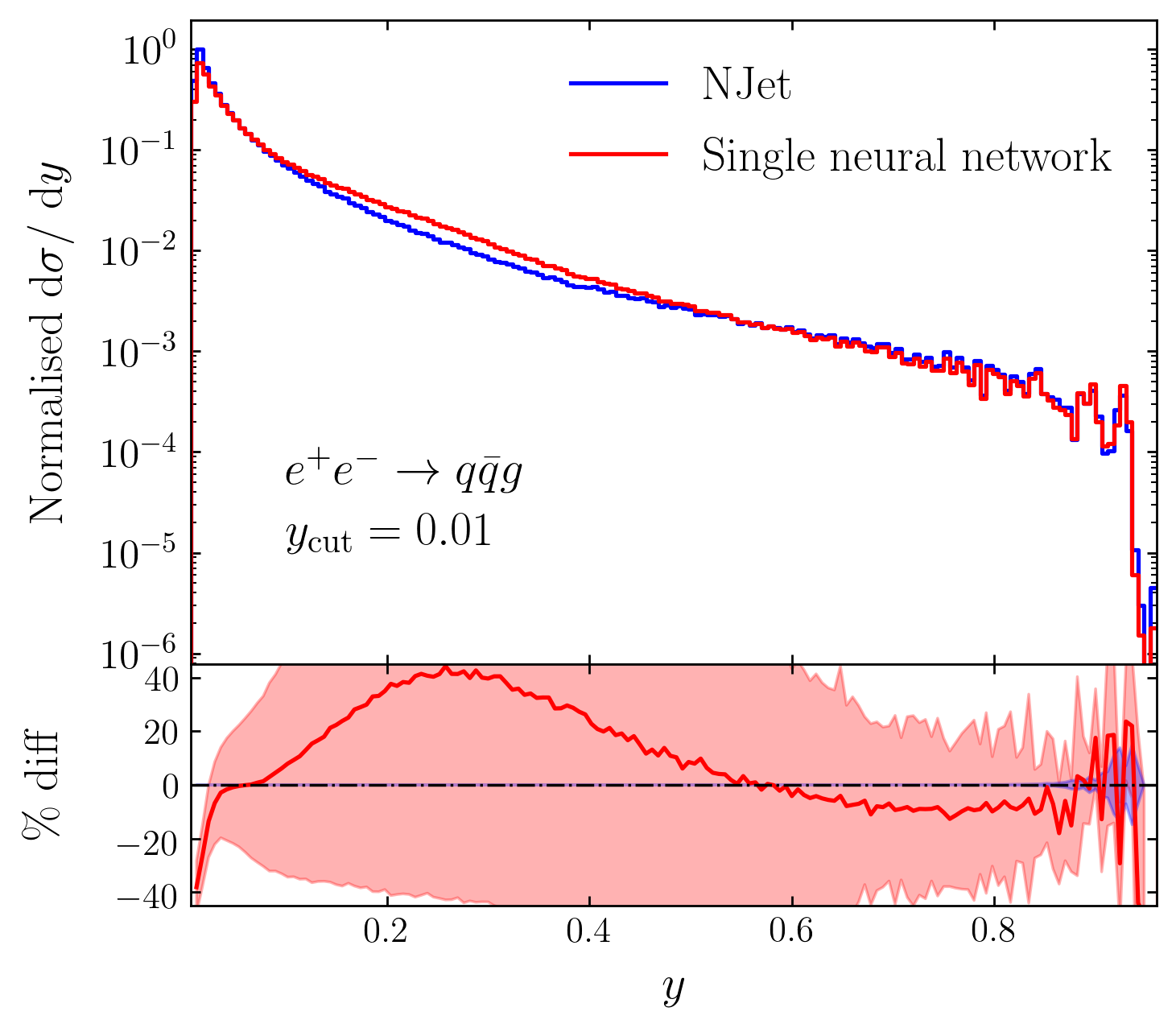}
\end{subfigure}%
\hfill
\begin{subfigure}{.49\textwidth}
    \centering
    \includegraphics[width=0.92\textwidth]{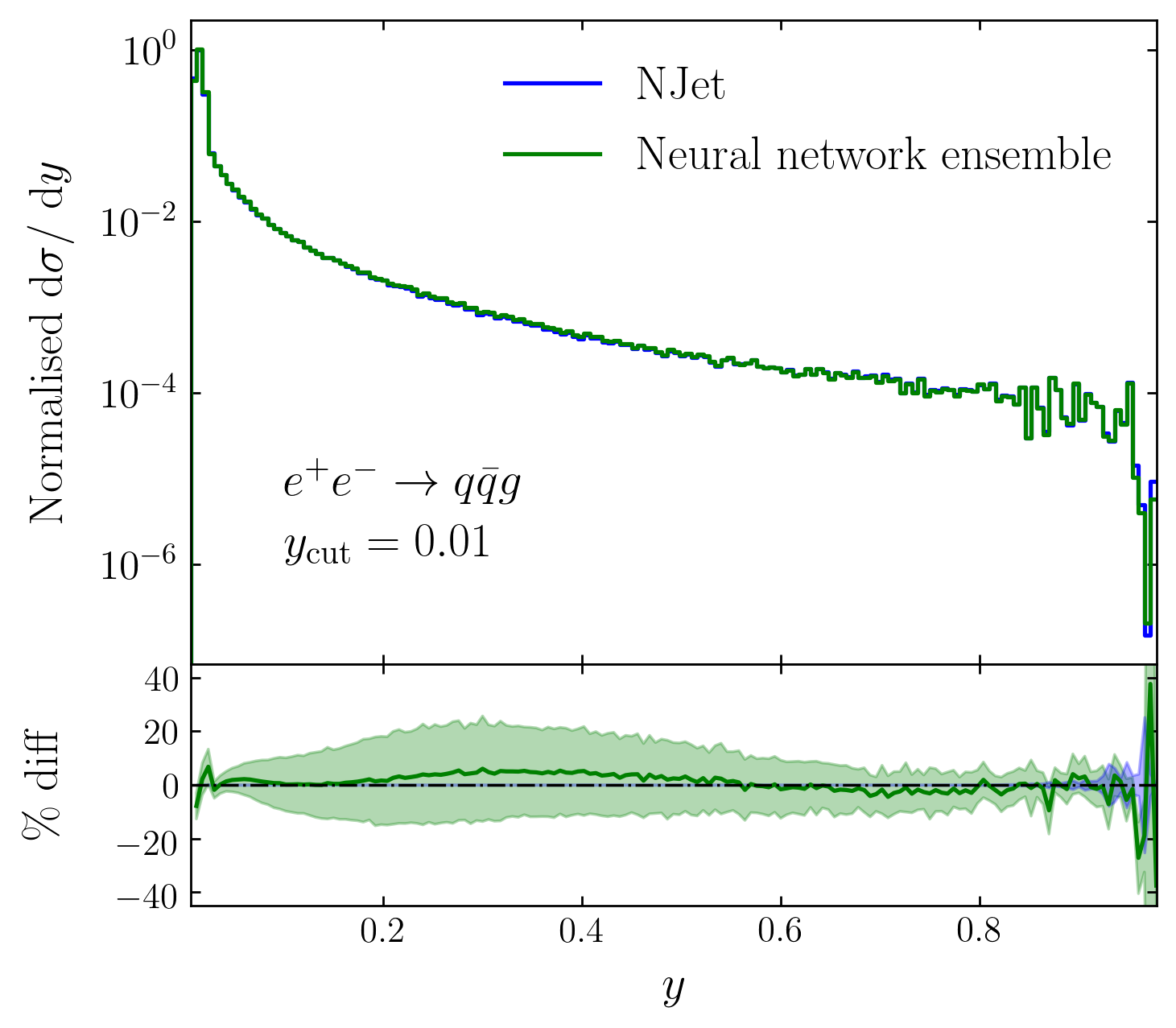}
\end{subfigure}

\centering
\begin{subfigure}{.49\textwidth}
    \includegraphics[width=0.92\textwidth]{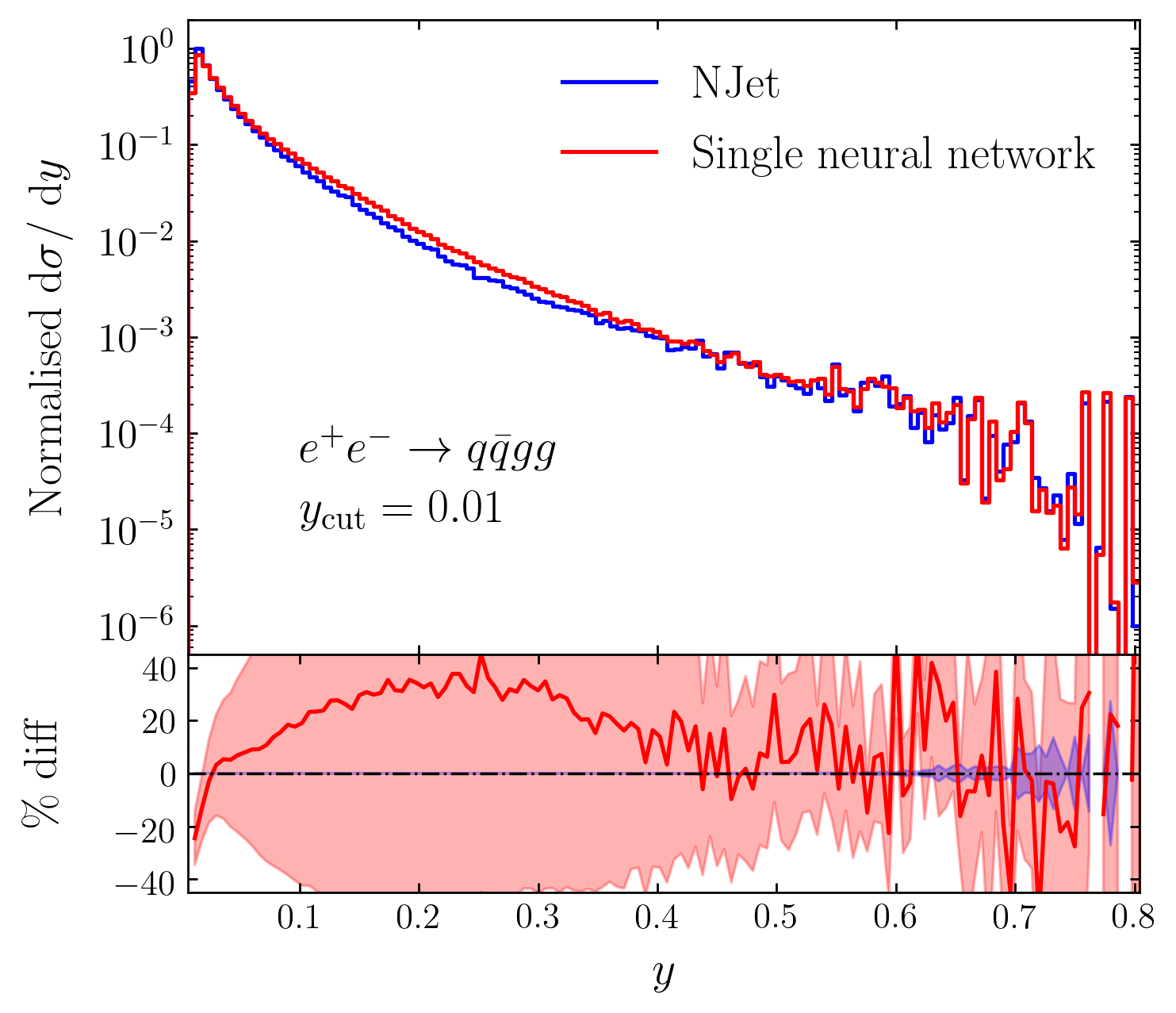}
\end{subfigure}%
\hfill
\begin{subfigure}{.49\textwidth}
    \centering
    \includegraphics[width=0.92\textwidth]{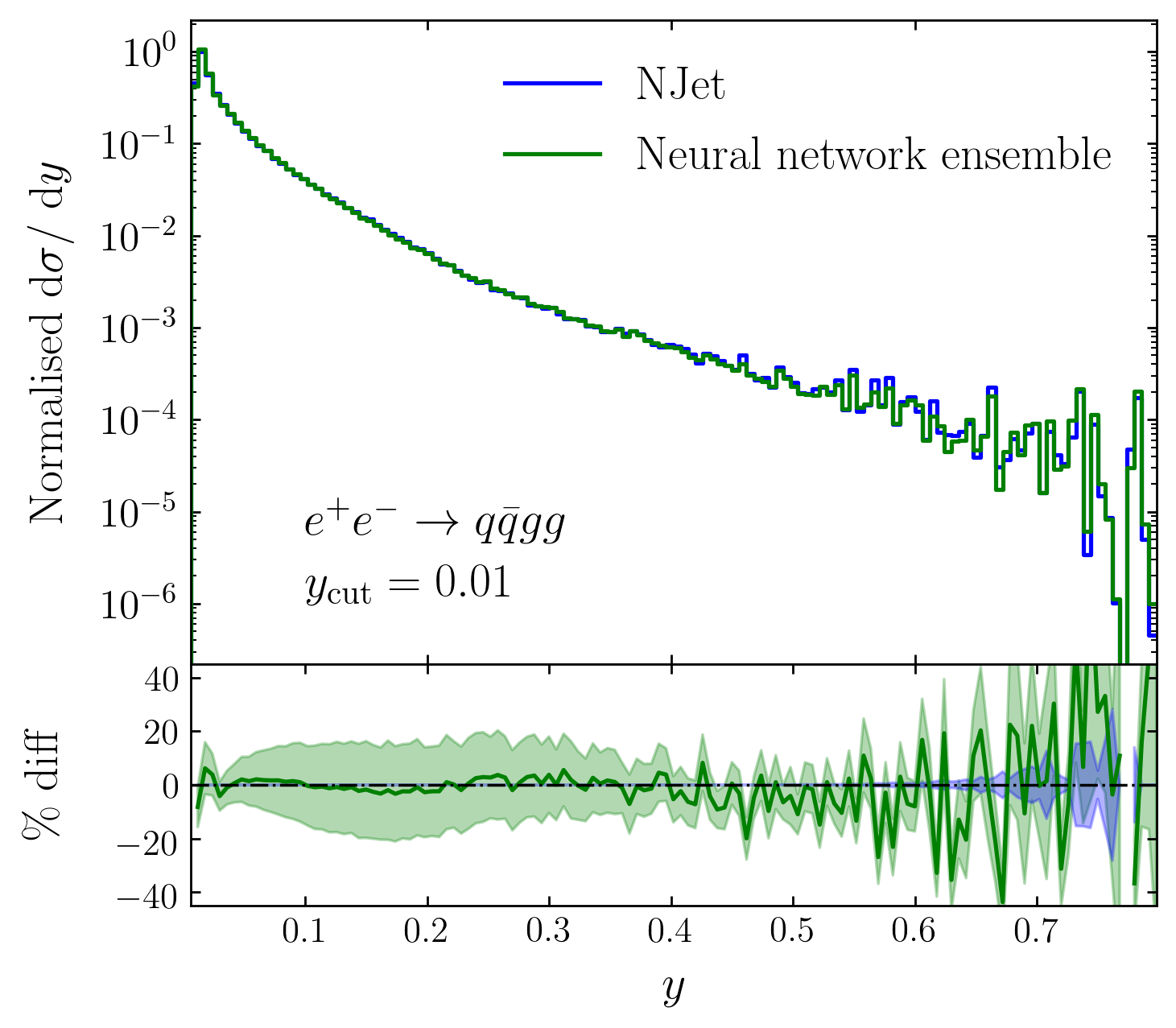}
\end{subfigure}

\centering
\begin{subfigure}{.49\textwidth}
    \includegraphics[width=0.92\textwidth]{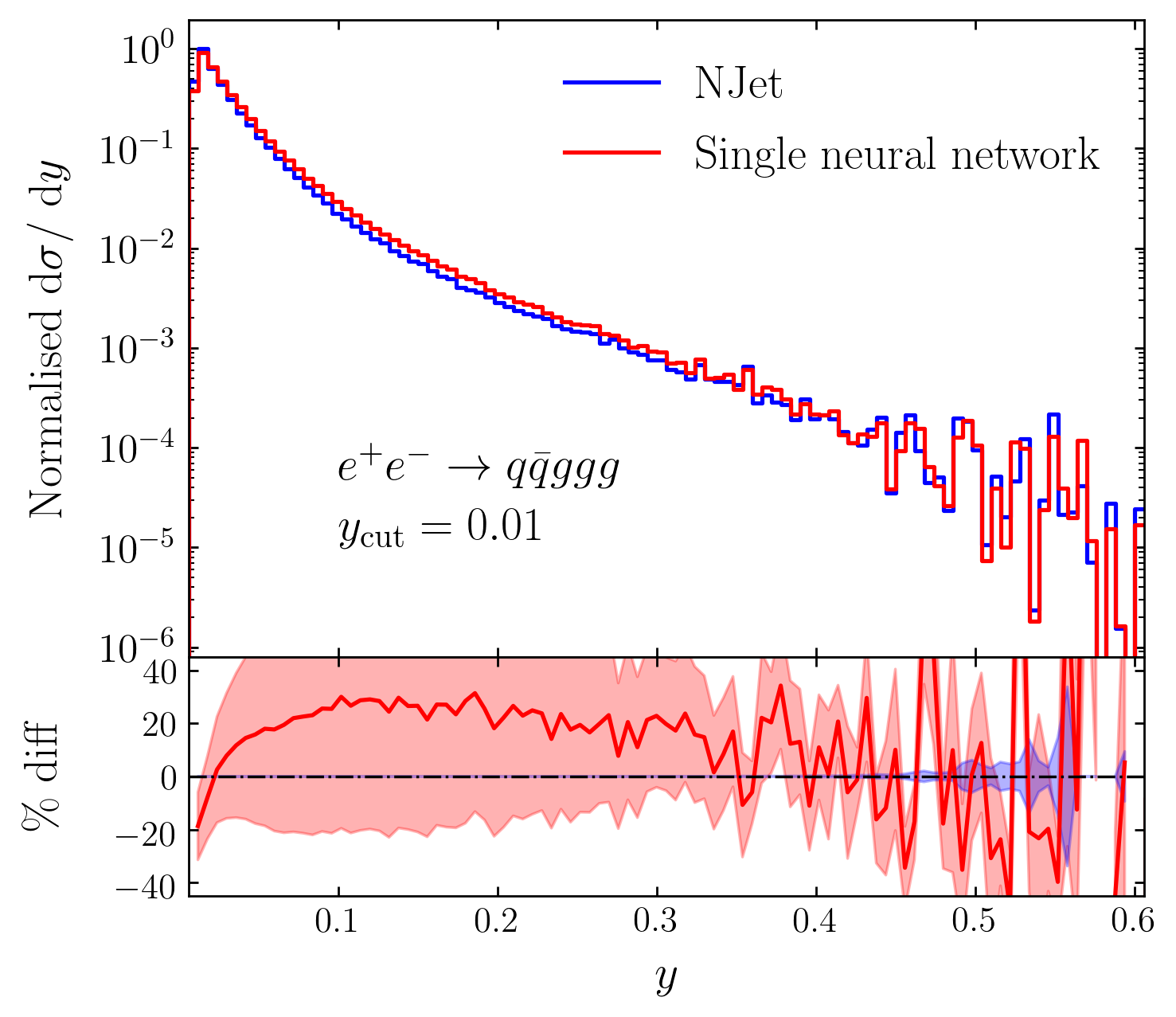}
\end{subfigure}%
\hfill
\begin{subfigure}{.49\textwidth}
    \centering
    \includegraphics[width=0.92\textwidth]{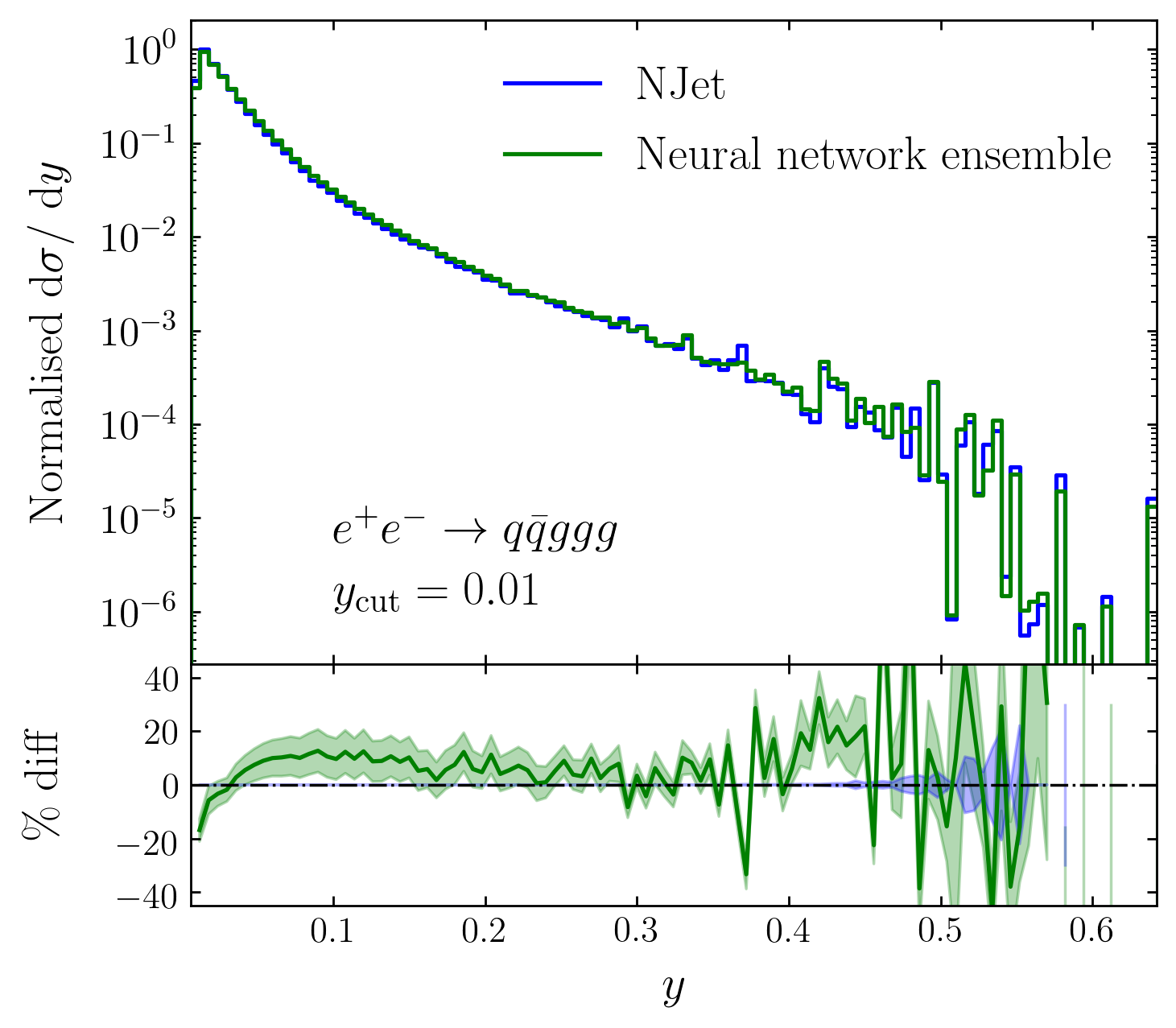}
\end{subfigure}

\centering
\begin{subfigure}{.49\textwidth}
    \includegraphics[width=0.92\textwidth]{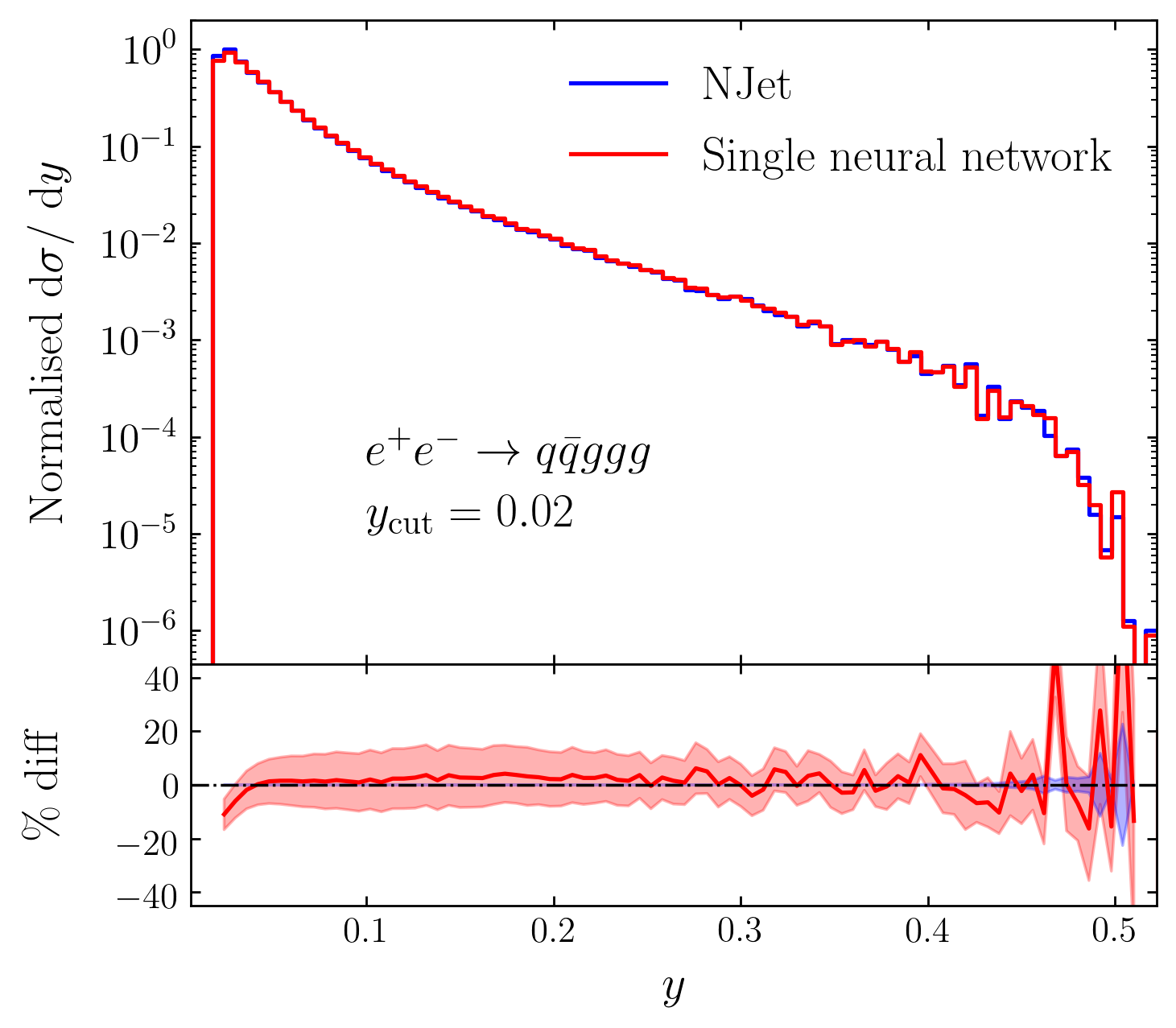}
\end{subfigure}%
\hfill
\begin{subfigure}{.49\textwidth}
    \centering
    \includegraphics[width=0.92\textwidth]{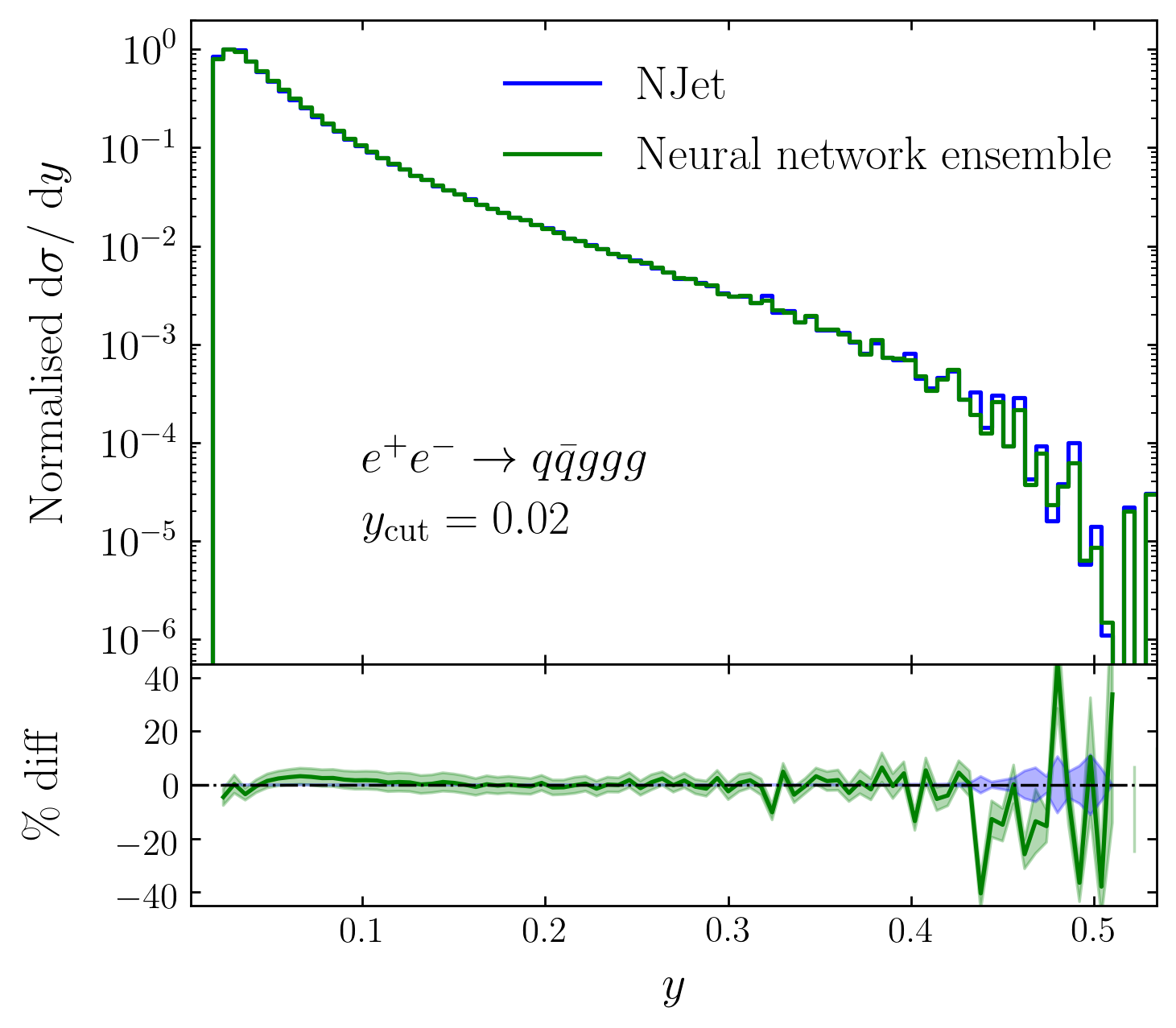}
\end{subfigure}
\caption{Comparison of a single neural network (left) vs. our ensemble approach (right) in estimating the differential cross-section against $y$, where $y$ is the minimum $y_{ij}$ as ordered by $p_T$. Data is normalised to the maximum \njet~bin value. Uncertainty bands denote 1 s.d. calculated over 20 trained models (red and green) and Monte Carlo error on the \njet~result (blue).}
\label{LO_diff_y}
\end{figure}

Let us first assume that the values of the cross-section calculated using the single network approach, $\sigma_s$ are normally distributed,\footnote{This is a reasonable assumption given that we would expect the uncertainty due to initialisation and dataset size to focus around a central mean value, with greater degrees of fluctuation becoming increasingly less likely. Additionally, any difference between the mean and the \njet~result would likely be systematic of the model architecture choice, sampling algorithm and other factors external to the uncertainty measured here, thus resulting in a symmetric distribution, up to an approximation.} i.e. $\sigma_s \sim \mathcal{N}(\mu_s, \zeta^2_s)$, where $\mu_s$ is the mean of the normal distribution and $\zeta_s$ is the standard deviation. Secondly, we note that in the case of the ensemble method the outputs of the networks trained over different partitions are first summed (c.f. Equation (\ref{FKS_cs_add})) giving:

\begin{equation}
\sigma_{\text{FKS}} = \sum_{p=1}^{N_{\text{max}}}\text{d}\sigma_p + \text{d}\sigma_{\text{non-div}},
\end{equation}

where $N_{\text{max}}$ is defined in Equation (\ref{N_max}),\footnote{In our implementation, for future process independence and coding simplicity we actually have $N_{\text{max}}+1$ pairs since we do not discard the $q\bar{q}$ pair. In the processes examined in this paper, this has the effect of splitting the non-divergent region into two parts although, given the ease with which the networks are able to learn this region, we do not find this causing an issue.} $\text{d}\sigma_p$ is the sum over all weighted matrix elements for a given FKS pair. Since we only partition the divergent region, $\mathcal{R}_{\text{div}}$, according to the FKS partition function, we add the differential cross-section over the non-divergent region, $\text{d}\sigma_{\text{non-div}}$. Given that the uncertainties in the individual networks making up the ensemble are expected to manifest themselves in a similar way to the single network approach, we may also assume that these are drawn from a normal distribution such that:

\begin{equation}\label{FKS_normal i_0}
\forall p\in\{1,...,N_{\text{max}}\}: \text{d}\sigma_p \sim \mathcal{N}(\mu_p, \zeta_p^2), \,\,\,\,\, \text{d}\sigma_{\text{non-div}} \sim \mathcal{N}(\mu_{\text{non-div}},\zeta_{\text{non-div}}^2),
\end{equation}
\begin{align}
\implies \sigma_{\text{FKS}} &\sim \sum_{p=1}^{N_{\text{max}}} \mathcal{N}(\mu_p, \zeta_p^2) + \mathcal{N}(\mu_{\text{non-div}}, \zeta_{\text{non-div}}^2) \label {FKS_normal i} \\
&\sim \mathcal{N}\left(\sum_{p=1}^{N_{\text{max}}}\mu_p, \sum_{p=1}^{N_{\text{max}}}\zeta_p^2\right) + \mathcal{N}(\mu_{\text{non-div}}, \zeta_{\text{non-div}}^2) \\
&\sim \mathcal{N}\left(\sum_{p=1}^{N_{\text{max}}}\mu_p + \mu_{\text{non-div}}, \sum_{p=1}^{N_{\text{max}}}\zeta_p^2 + \zeta_{\text{non-div}}^2\right)\\
& := \mathcal{N}(\mu_{\text{FKS}}, \zeta_{\text{FKS}}^2). \label {FKS_normal f}
\end{align}

Since the uncertainties in the ensemble method are smaller than those found when using the single network approach:

\begin{equation}
\zeta_{\text{FKS}}^2 < \zeta_s^2
\end{equation}
\begin{equation}\label{zeta p,s comparison}
\implies \zeta_p^2 < \zeta_s^2,  \forall p\in \{1,...,N_{\text{max}}\}\,\, \text{and}\,\, \zeta_{\text{non-div}}^2 < \zeta_s^2.
\end{equation}
\newpage

From Equation (\ref{zeta p,s comparison}) we see that not only does the ensemble of models have a reduced uncertainty in comparison to the single model approach, but that each individual model making up the ensemble also has a reduced uncertainty, thus supporting the claim that by using the ensemble method, the models learning the divergent structure are more certain about what they are learning and less sensitive to both model initialisation and dataset size. 

The overall accuracy of the ensemble approach, combined with the implications of Equation (\ref{zeta p,s comparison}), demonstrates that we are learning the divergent structure of the amplitude sufficiently well. As discussed in Section \ref{sec:errors_analysis}, it should be noted that Figure \ref{LO_cs} does not show the performance of a single model, but rather the average of 20 trained models with their equivalent standard error. Although one does not have to train this many models to still get a good approximation, in Section \ref{NLO/LO results} we will see that training additional models is computationally cheap and thus not a large hinderance. For comparison, the performance of training just a single model is shown in Figure \ref{LO_diff_y} as discussed below.

Figure \ref{LO_diff_y} shows the differential cross-section of the $y_{ij}$ distribution of the two softest jets as ordered by $p_T$. Here, we still plot the mean of the 20 trained models, but now state the standard deviation from the mean to more clearly show the spread of model performance and the effect of only training a single model. These differential distributions were chosen as they highlight the performance of the models in hard-to-sample regions of phase-space, in particular some of the regions we would expect the FKS partition function to assist with learning. Indeed, we see a significant improvement in comparison to the performance of the single neural network approach when using our ensemble method both in overall per-bin accuracy and stability. In addition, the ensemble method also produces narrower uncertainty bands than the single network approach, thus demonstrating its higher confidence in these regions. While this confidence is seen to be slightly misplaced in the case of the residual peak of the 5-jet plot at $y_{cut} = 0.01$, we see the harsher cut correcting for this and producing good agreement between the \njet~and ensemble results. Similar reasoning as given in Equations (\ref{FKS_normal i_0} - \ref{zeta p,s comparison}) can be applied to the per-bin uncertainty differences between the single and ensemble network approaches.

Overall, the ensemble of networks is shown to produce more accurate and reliable results in LO approximations than the single network approach. While it can be argued that there is greater computational expense in training multiple networks, given the very low cost of network training in comparison to the data generation time this is considered to be negligible, particularly at higher orders (see Section \ref{NLO/LO results} for more details).

\subsection{Virtual Approximations at NLO \label{NLO/LO results}}

When approximating the k-factor, the infrared singularities present in the previous examples have been normalised. This normalisation regulates the number of large divergences in phase-space, allowing the network to focus more on learning the loop-induced divergences. Additionally, although the FKS method is especially useful for isolating soft and collinear divergences at LO, given the presence of $\text{log}(s_{ij})$ terms in the virtual corrections we still expect to see improvements by using the ensemble method when approximating the k-factor.

\begin{figure}
\centering
\begin{subfigure}{.49\textwidth}
    \includegraphics[width=\textwidth]{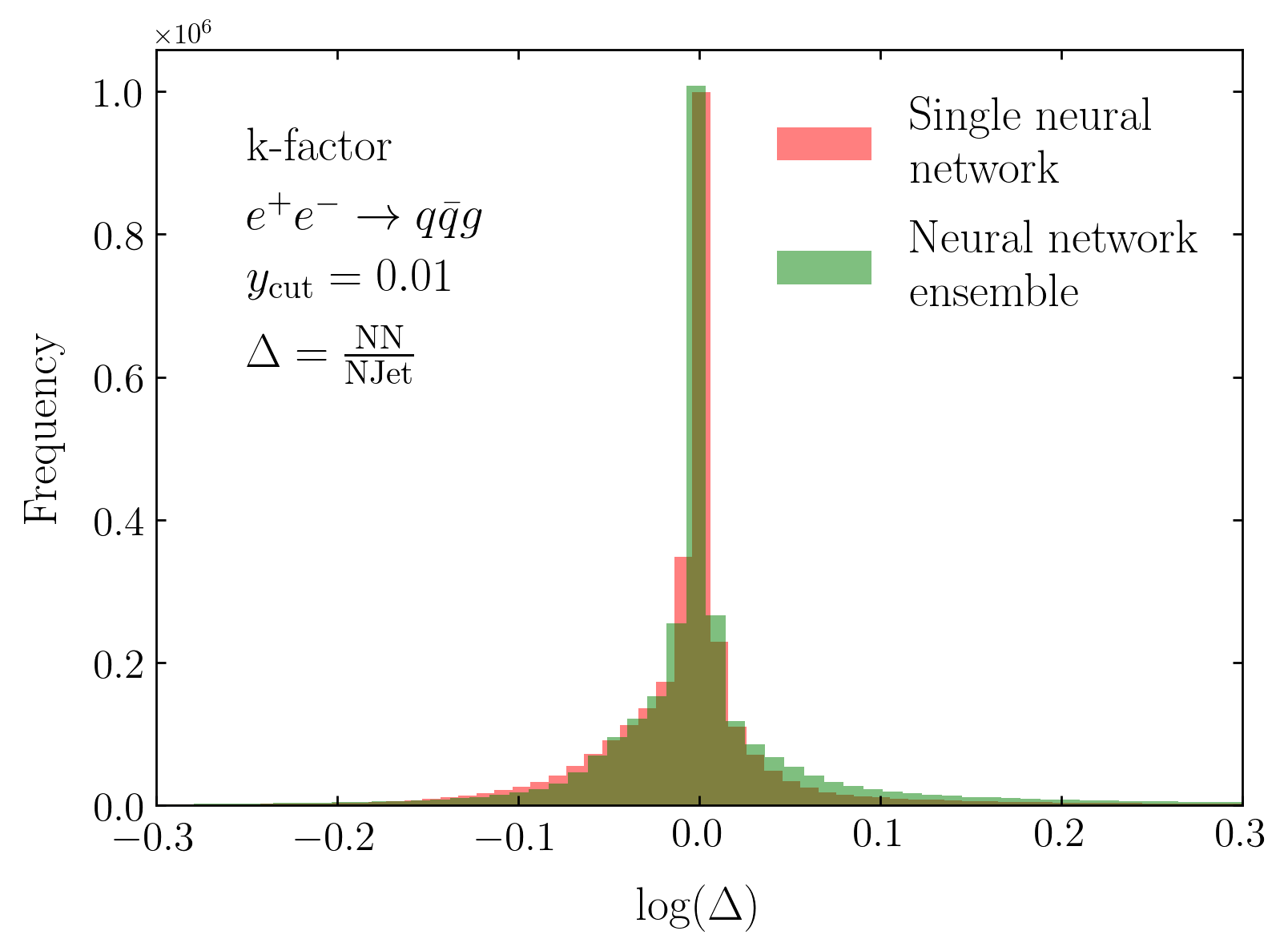}
\end{subfigure}%
\hfill
\begin{subfigure}{.49\textwidth}
    \centering
    \includegraphics[width=\textwidth]{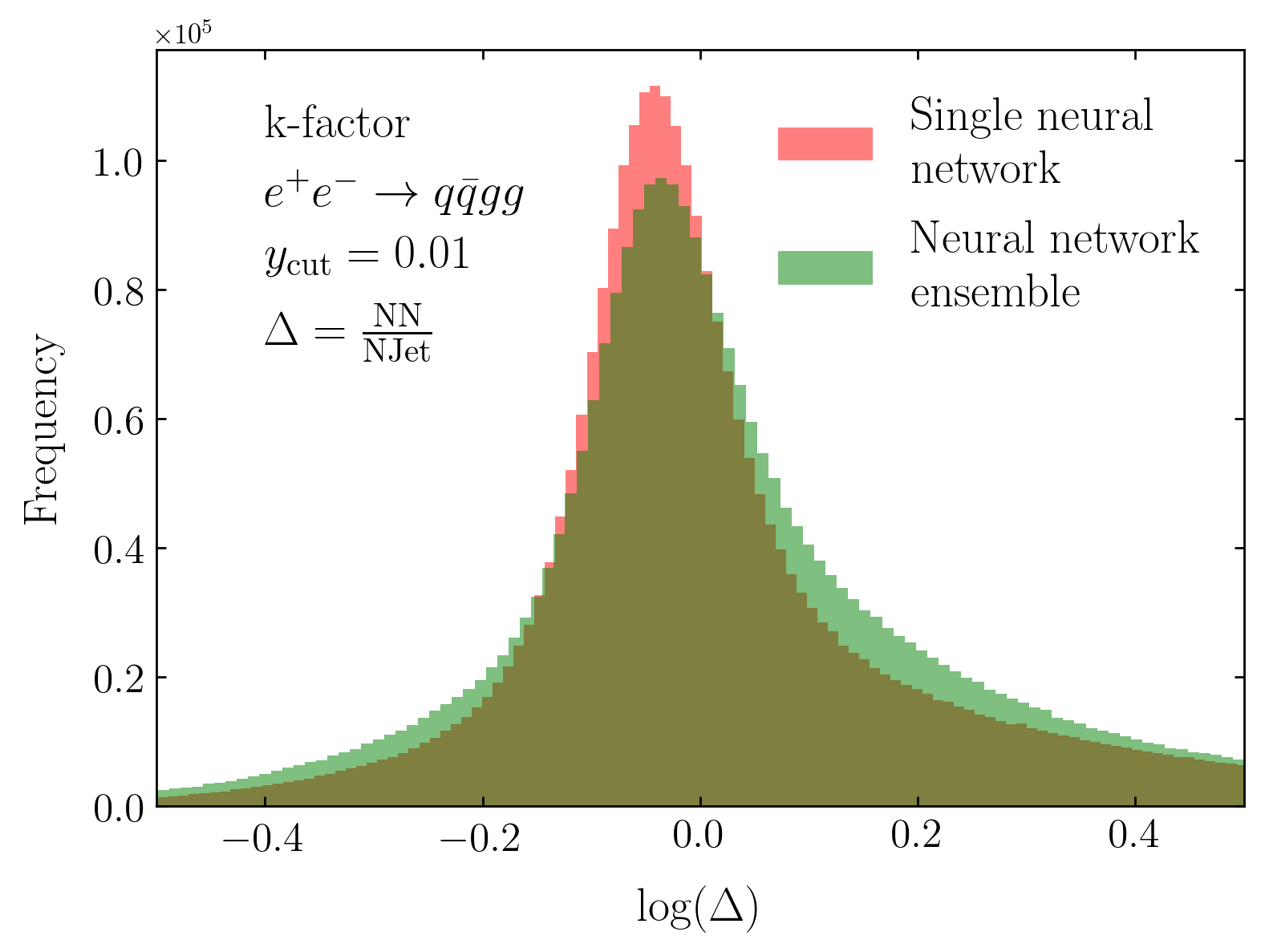}
\end{subfigure}
\caption{k-factor output of a single neural network (red) and our ensemble approach (green) compared to the \njet~calculation. Outputs are taken as the average over 20 trained models or ensembles.}
\label{NLO_errors}
\end{figure}

As in the LO case, in Figure \ref{NLO_errors} we plot the error distributions for the single and ensemble cases by comparing the network outputs to the \njet~calculations at the per-point level. In the 3 and 4-jet cases we see that both methods perform relatively similarly, with the single network appearing to be slightly better in the case of 4-jets. However, it should again be noted that these plots do not contain information about the network uncertainty and so should not be interpreted as the sole measure of performance.


\begin{figure}
\centering
\begin{subfigure}{.49\textwidth}
    \includegraphics[width=\textwidth]{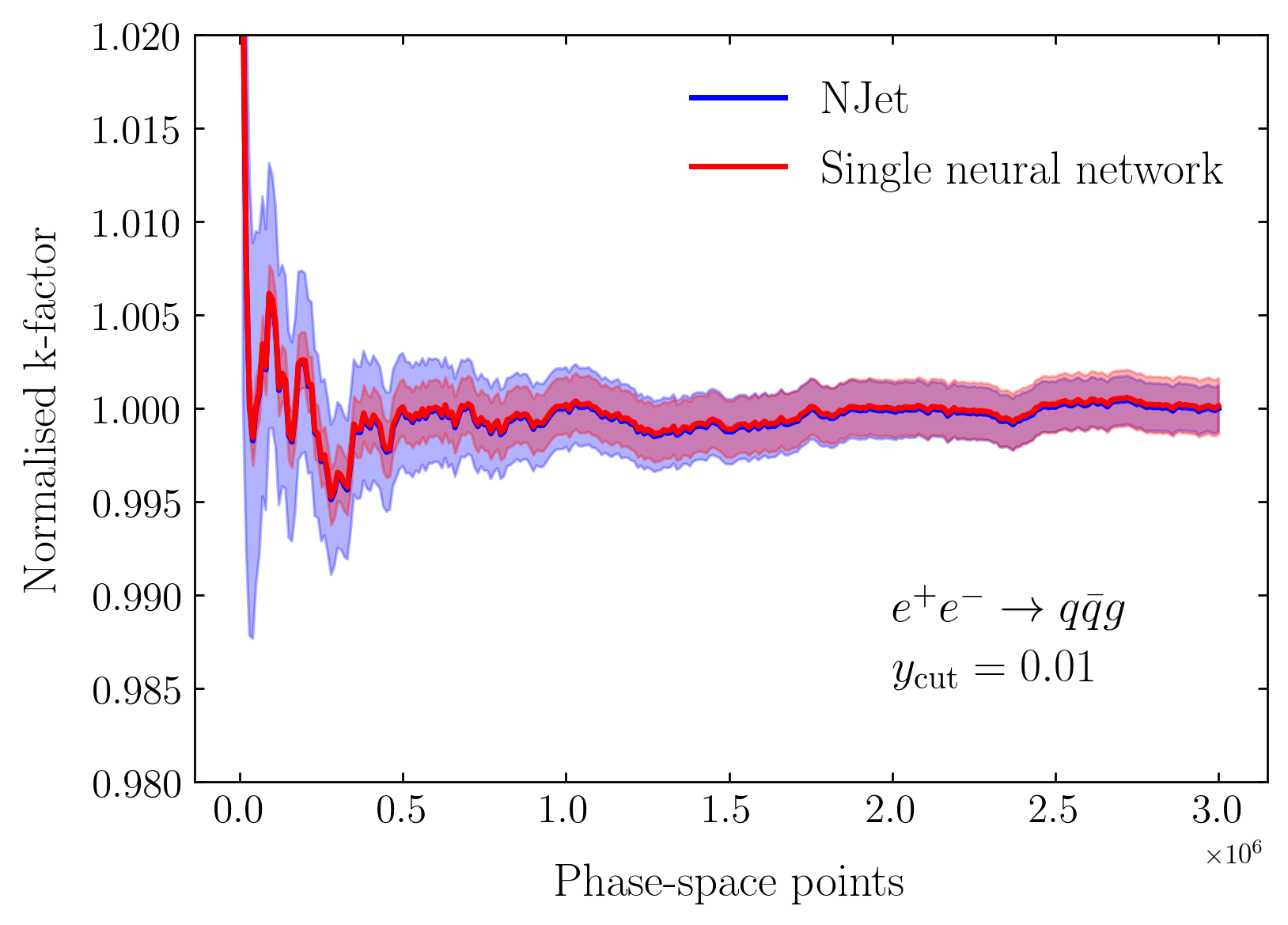}
\end{subfigure}%
\hfill
\begin{subfigure}{.49\textwidth}
    \centering
    \includegraphics[width=\textwidth]{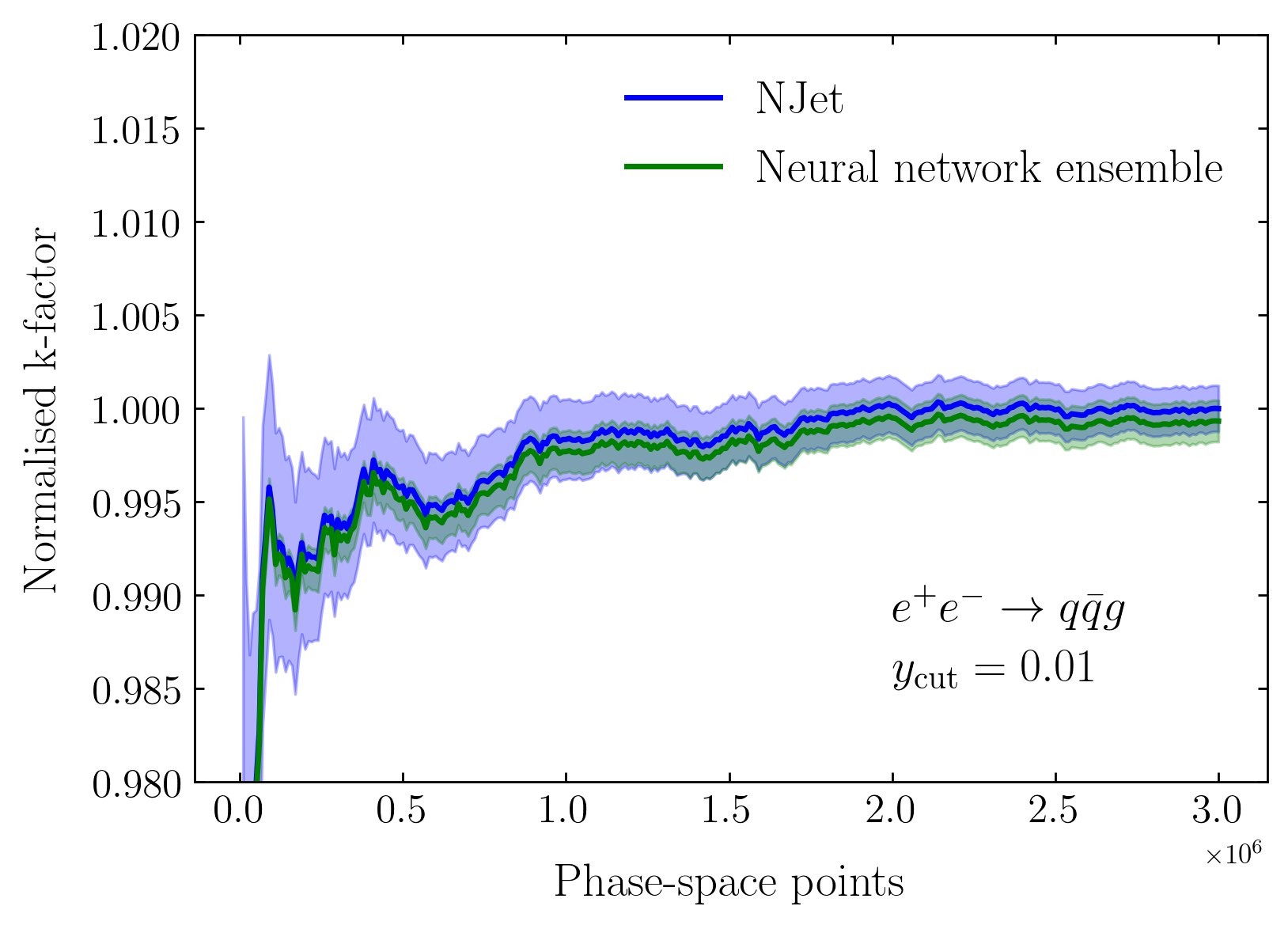}
\end{subfigure}

\begin{subfigure}{.49\textwidth}
    \includegraphics[width=\textwidth]{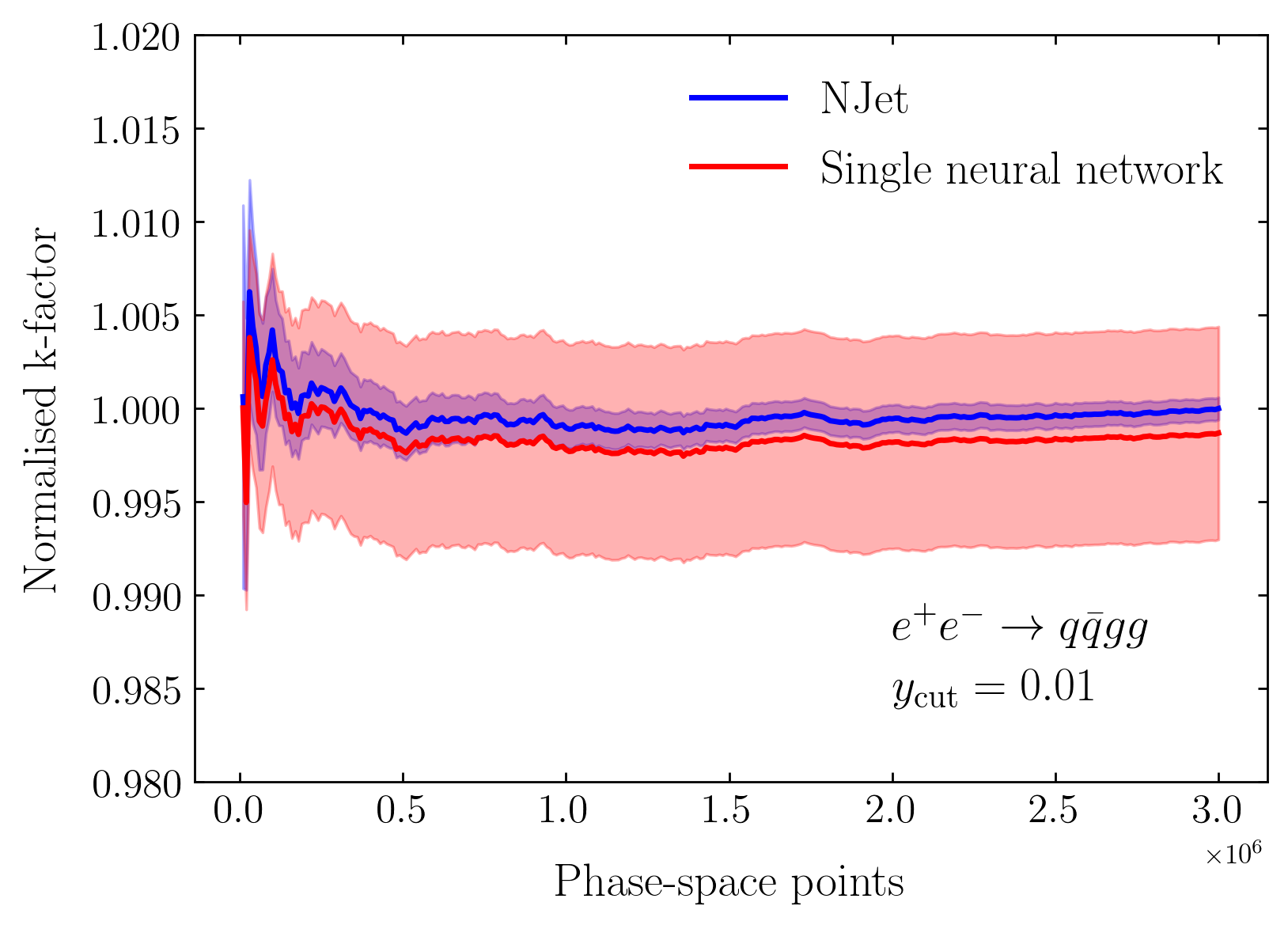}
\end{subfigure}%
\hfill
\begin{subfigure}{.49\textwidth}
    \centering
    \includegraphics[width=\textwidth]{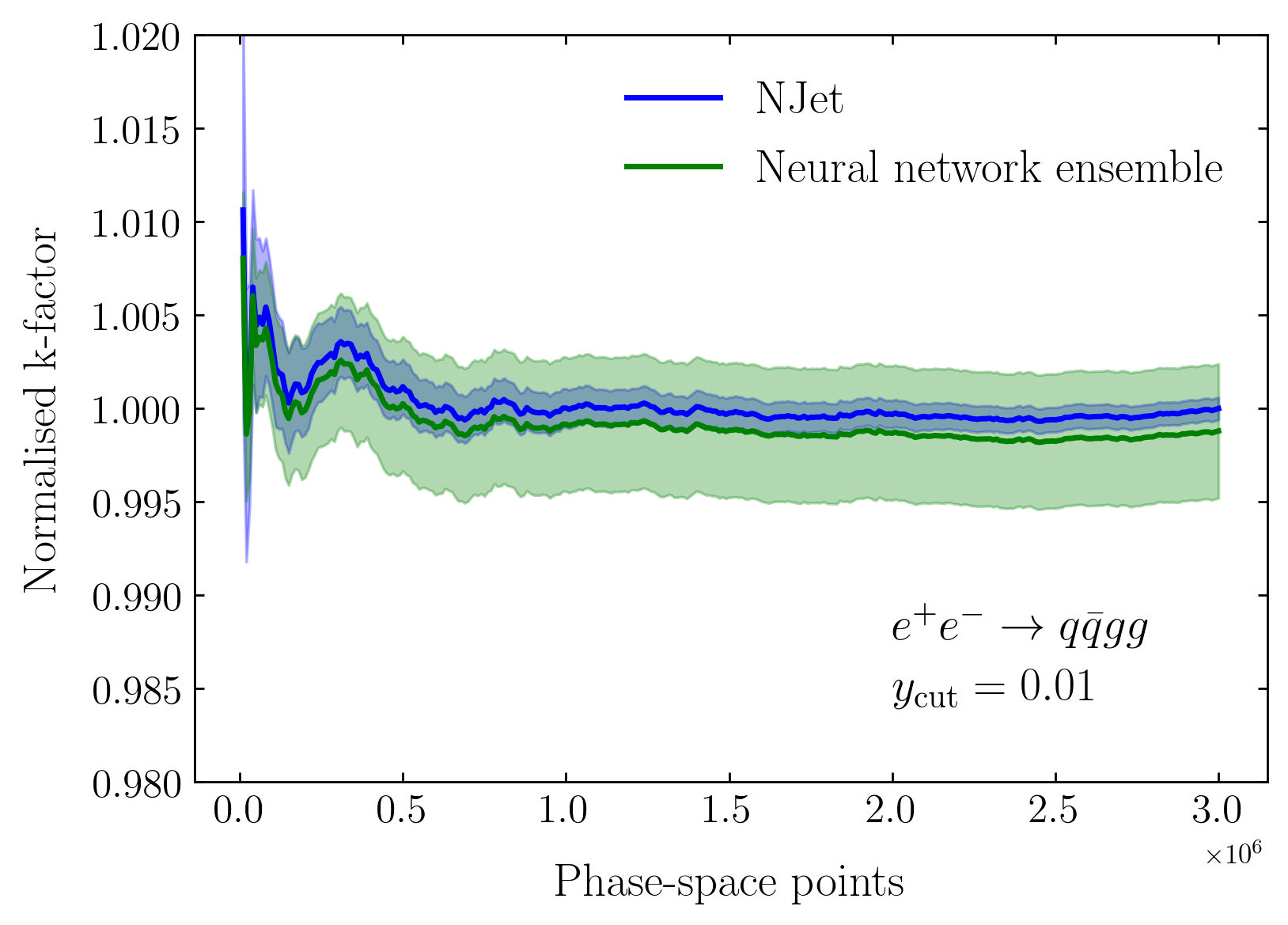}
\end{subfigure}
\caption{Comparison of a single neural network (left) vs. our ensemble approach (right) in estimating the normalised NLO/LO k-factors. Uncertainty bands denote the standard error on the mean calculated over 20 trained models (red and green) and Monte Carlo error on the \njet~result (blue).}
\label{k_factors}
\end{figure}


\begin{figure}
\centering
\begin{subfigure}{.49\textwidth}
    \includegraphics[width=\textwidth]{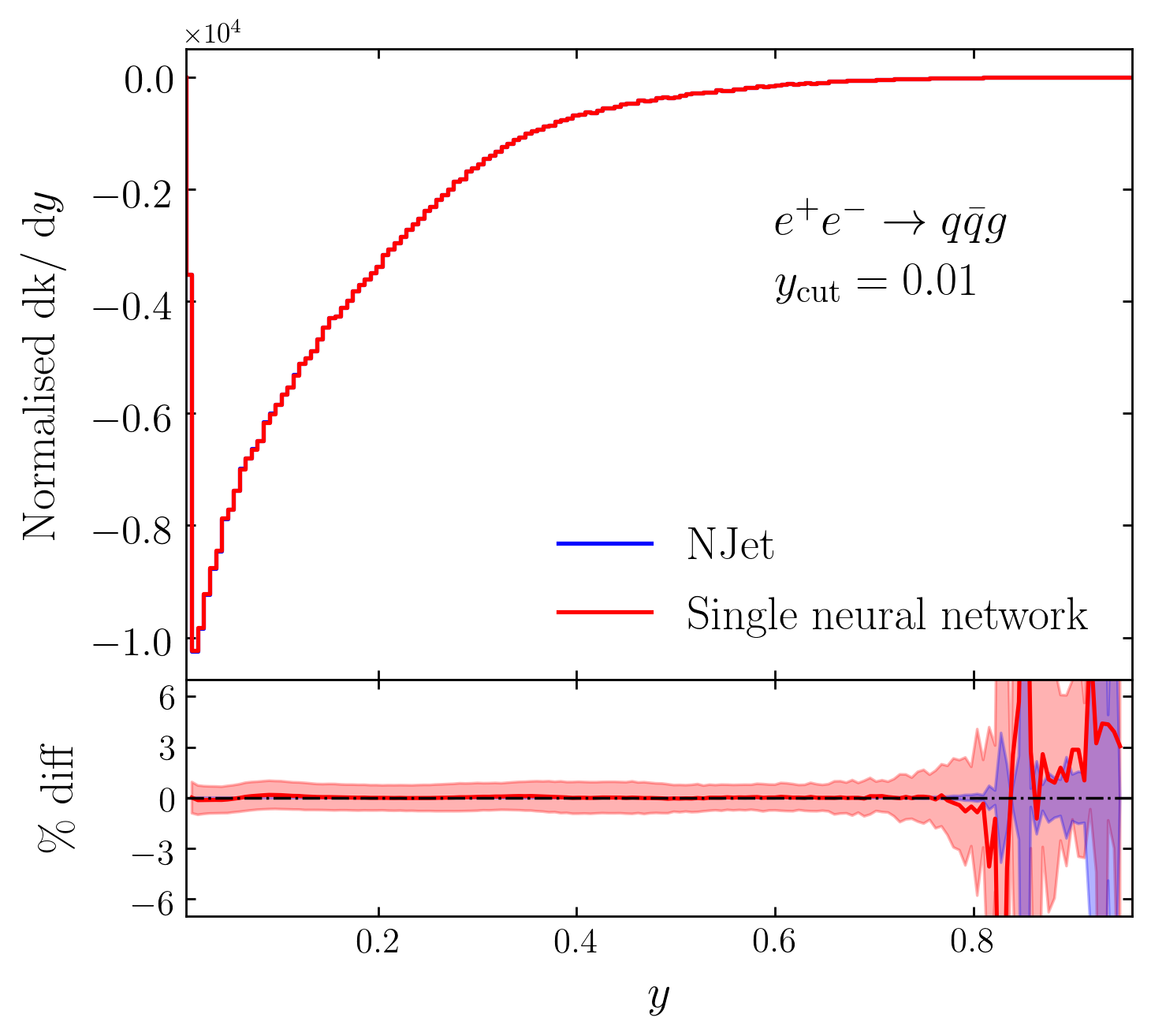}
\end{subfigure}%
\hfill
\begin{subfigure}{.49\textwidth}
    \centering
    \includegraphics[width=\textwidth]{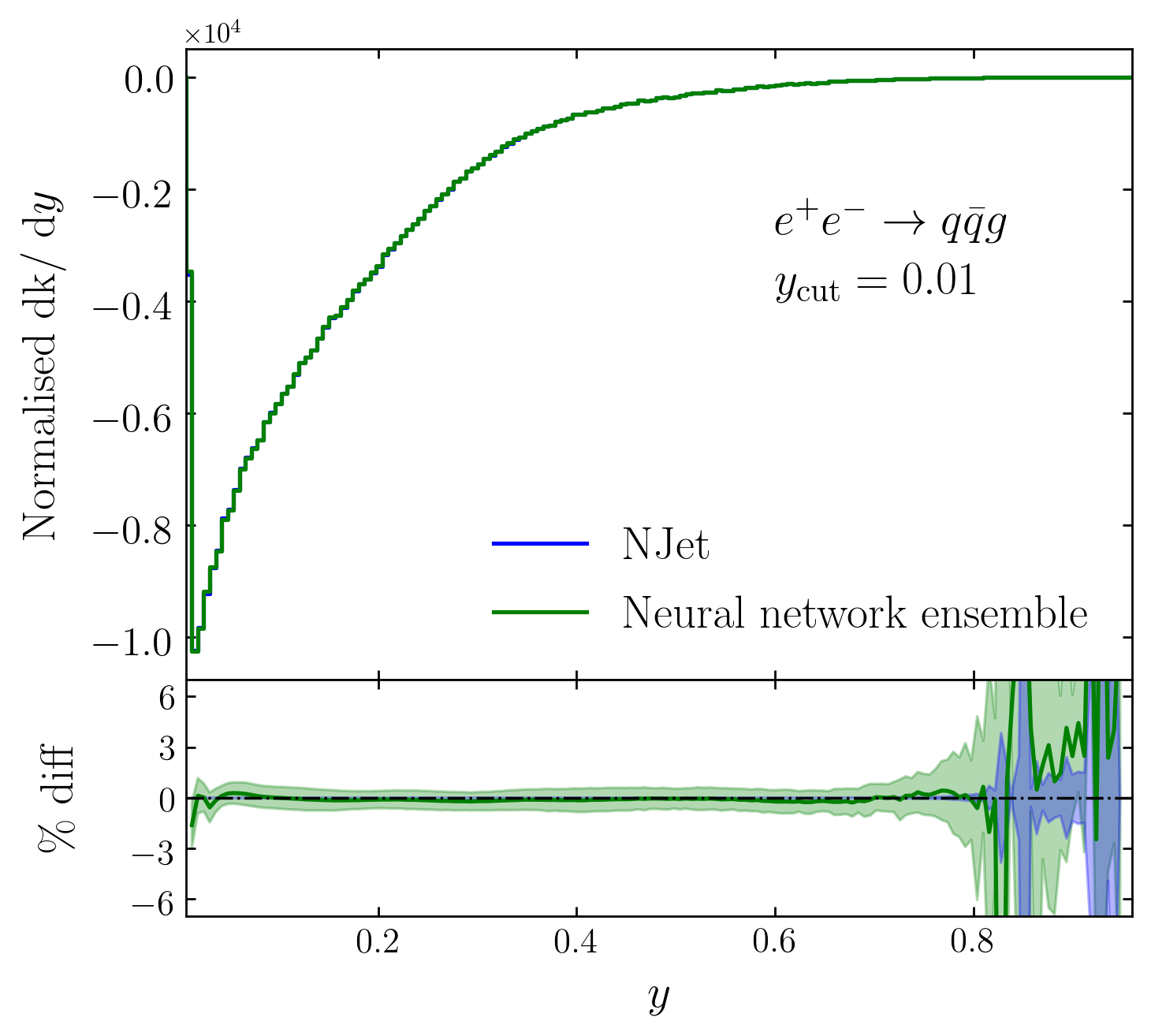}
\end{subfigure}

\centering
\begin{subfigure}{.49\textwidth}
    \includegraphics[width=\textwidth]{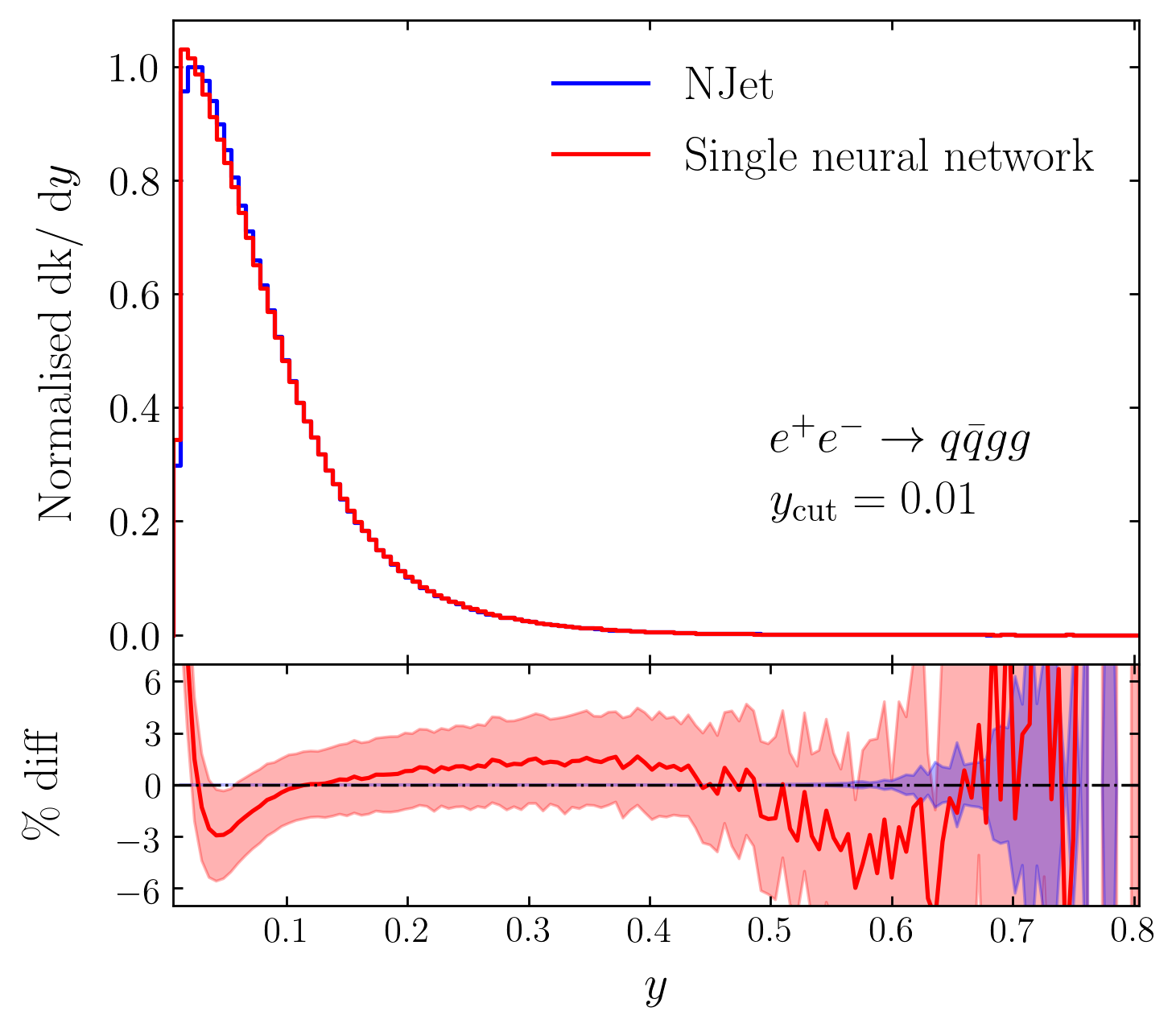}
\end{subfigure}%
\hfill
\begin{subfigure}{.49\textwidth}
    \centering
    \includegraphics[width=\textwidth]{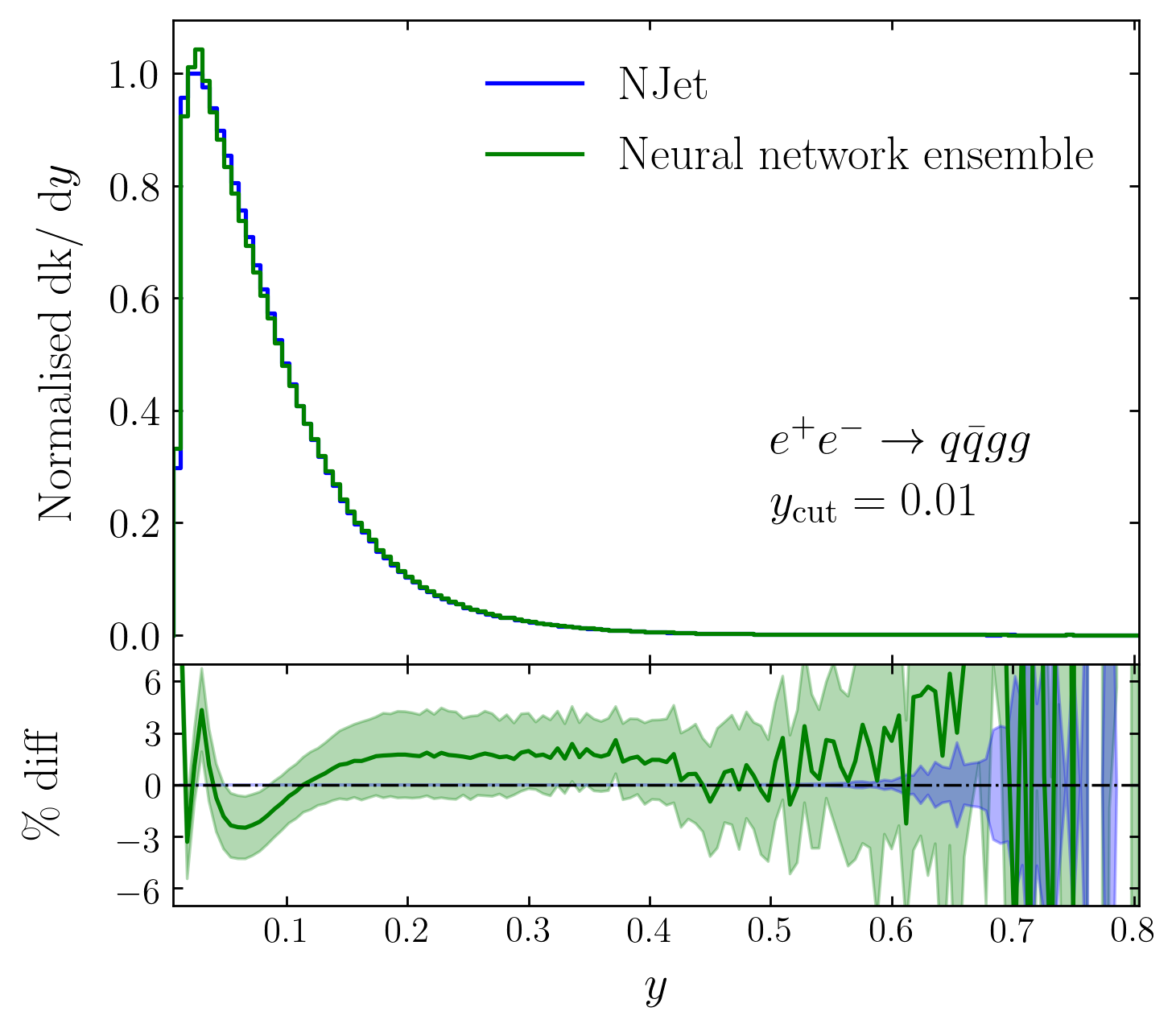}
\end{subfigure}
\caption{Comparison  of a single neural network (left) vs. our ensemble approach (right) in estimating the differential NLO/LO k-factors against $y$, where $y$ is the minimum $y_{ij}$ as ordered by $p_T$. Data is normalised to the maximum \njet~bin value. Uncertainty bands denote the s.d. calculated over 20 trained models (red and green) and Monte Carlo error on the \njet~result (blue).}
\label{diff_k_factors_y}
\end{figure}


\begin{table}
\begin{center}
    \begin{tabular}{|c|c|c|c|c|}
    \hline
    &\multicolumn{2}{|c|}{10k training, 1M inference}& \multicolumn{2}{|c|}{100k training, 1M inference} \\ \hline
    Jets & \njet & NN ensemble &  \njet & NN ensemble\\
    \hline
    3 & 13.2 hours &  0.15 hours &  13.2 hours & 1.32 hours \\
    4 & 194 hours &  1.97 hours & 194 hours & 19.4 hours \\
    5 & $6.39\times10^3$ hours & 63.9 hours & $6.39\times10^3$ hours & 639 hours \\ \hline

    \end{tabular}
\end{center}
    \caption{Time required for k-factor calculation at different multiplicities requiring 1M points while training on 10k and 100k points. These results assume all calculations take place on a single CPU core. Note that the \njet~times remain the same as we assume that the training points form part of the inference (see later in this section for more details). Time is quoted to at most 2 decimal places and at 3 significant figures where possible. \label{comp_comparison}}
\end{table}


\begin{figure}


\centering
\begin{subfigure}{.49\textwidth}
    \includegraphics[width=\textwidth]{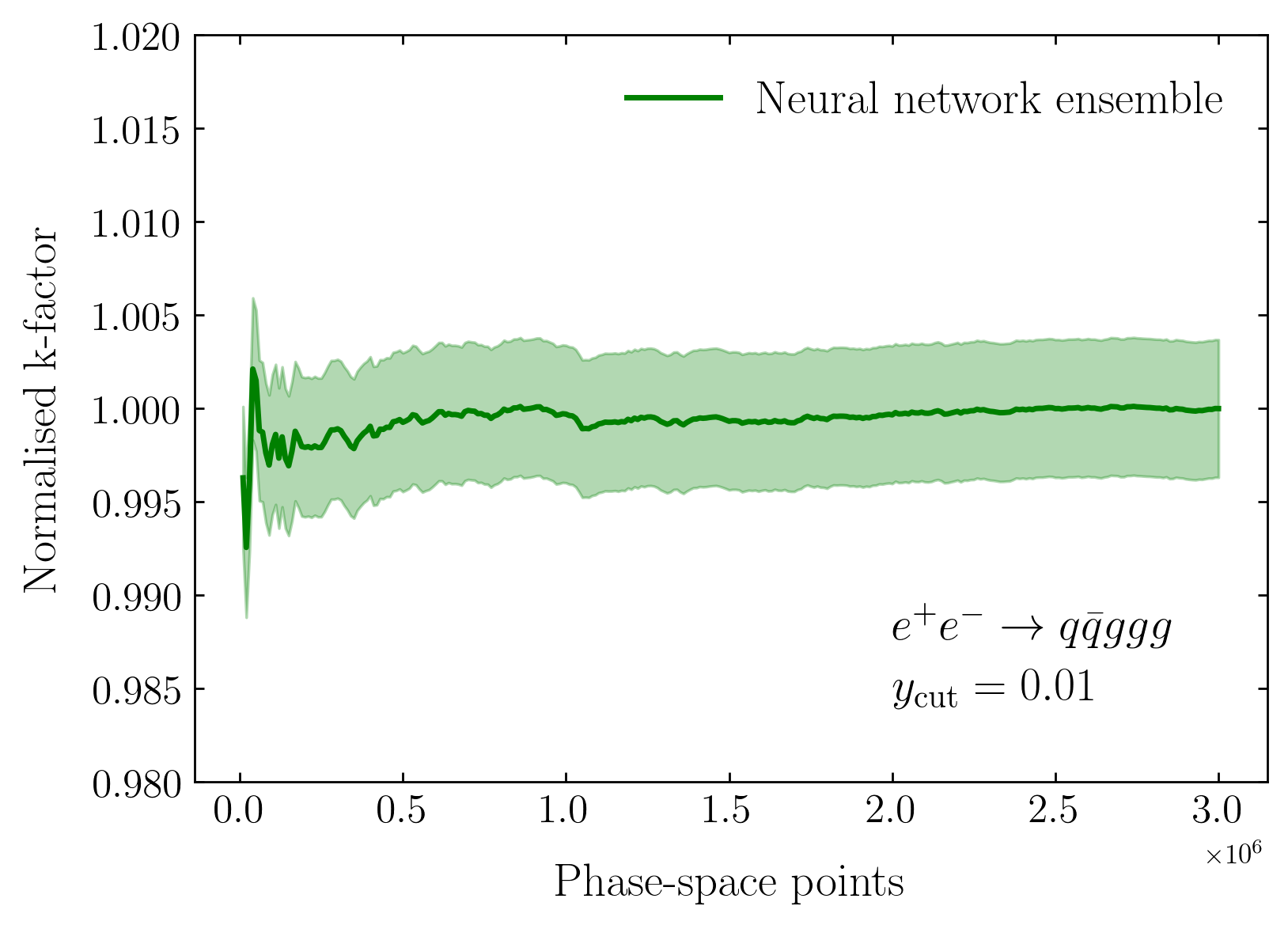}
\end{subfigure}%
\hfill
\begin{subfigure}{.49\textwidth}
    \centering
    \includegraphics[width=\textwidth]{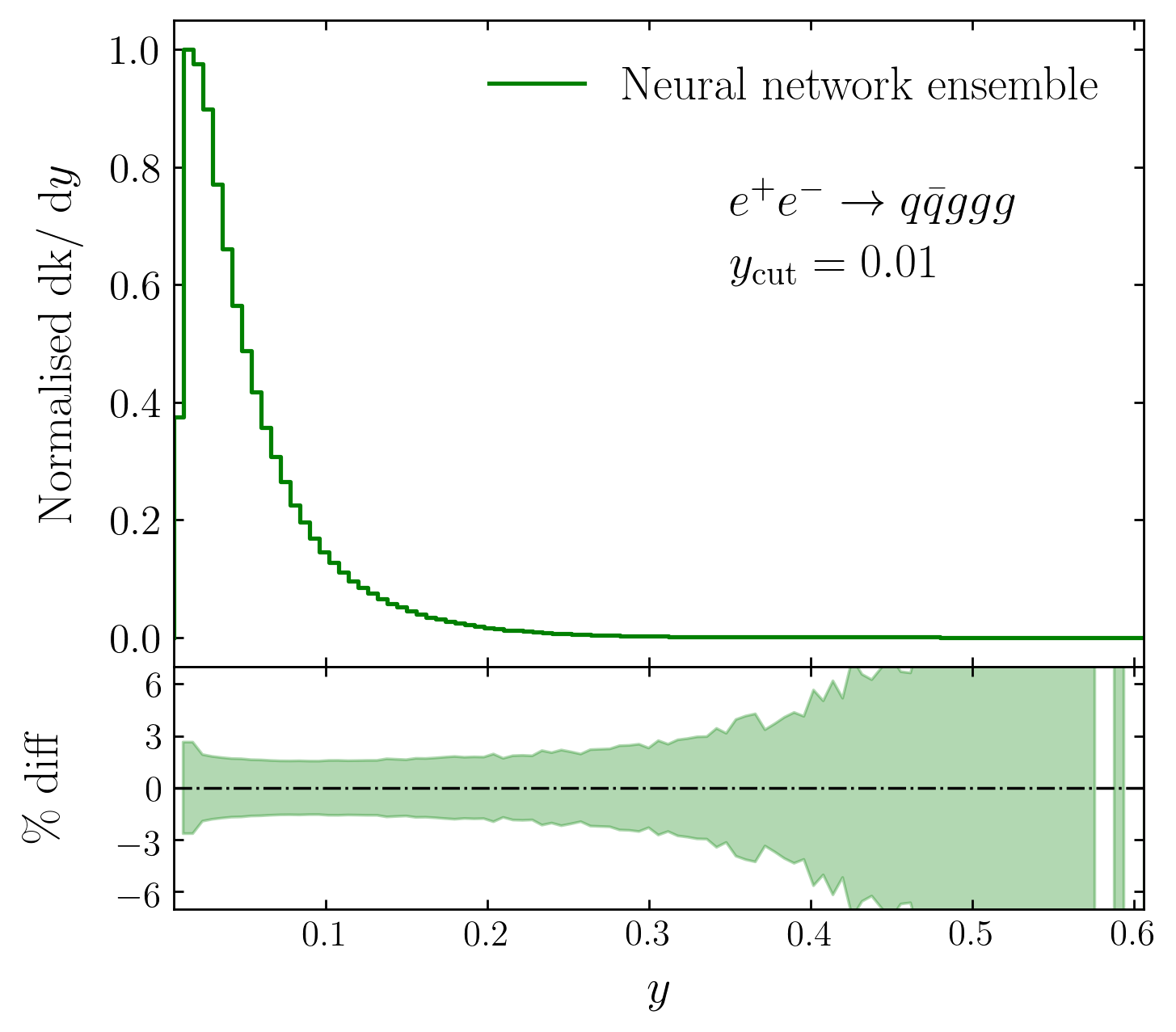}
\end{subfigure}
\caption{Normalised NLO/LO k-factor and differential k-factor against $y$, where $y$ is the minimum $y_{ij}$ as ordered by $p_T$, at 5 jets using just a the neural network ensemble. Data is in the differential plot is normalised to the maximum network output value. Uncertainty bands denote 1 standard error in the k-factor plot and 1 standard deviation in the differential plot, both as a percentage of the mean calculated over 20 trained models.}
\label{NLO_5_jets}
\end{figure}

In Figure \ref{k_factors} we see that both the naive single network approach and the ensemble method approximate the k-factor to within Monte Carlo error at 3-jets, and are within the percent level at 4-jets. Although either methodology would be suitable for use, the ensemble of networks requires little more computational time in comparison to the single network model, while producing narrower uncertainty bands. For robustness at higher multiplicity, the ensemble method remains the more optimal method. 

A comparison between the computational speed of different methods of k-factor computation and calculation can be found in Table \ref{comp_comparison}. Here we see a dramatic speed up when using the network approximation as opposed to current numerical methods, with the dominant time saving coming from the reduction of the number of matrix elements having to be explicitly calculated using \njet~(i.e. in the case of training on 100k points and inferring on 1M at high multiplicity the speed up is $\mathcal{O}(10)$). Moreover, the assertion that the ensemble method is not significantly more expensive than the naive approach can be verified. It should be noted that by only training on 10k points we may have an unacceptable accuracy when compared to the 100k results. The results presented in the table are therefore designed to demonstrate the computational time required for network training in comparison to the \njet~calculation, as opposed to providing guidelines on how many training points to use.


As in Section \ref{LO results}, we plot the differential k-factors of the $y$ distribution of the two softest jets as ordered by $p_T$. In Figure \ref{diff_k_factors_y} we see that both the single network, and ensemble approach, model the data well, with uncertainties at the 1-2 \% level. As before, the ensemble method provides us with slightly narrower uncertainty bands in both the 3-jet and 4-jet cases. Additionally, although neither the single network, nor ensemble approach approximate the peak in the 4-jet distribution exactly, the peak location is more accurately approximated by the ensemble approach with only a single bin at the peak being significantly ill-approximated. While we do not necessarily see a dramatic improvement in using the ensemble approach, given that the additional training time required is negligible in comparison to the data generation, we still see it as a viable and beneficial method to use for k-factor approximation. It should be noted that similar reasoning as given in Equations (\ref{FKS_normal i_0} - \ref{zeta p,s comparison}) can again be applied to the k-factor and per-bin uncertainty differences between the single and ensemble network approaches at NLO.

Finally, in the case of 5-jets we demonstrate our methodology as it may be used in practice. In Figure \ref{NLO_5_jets} we show how one may predict on a set of points with no known \njet~results for testing, while understanding the associated neural network errors. From these plots we clearly see that the ensemble method has associated errors only at the level of 2\% in the cross-section, with larger uncertainties in the regions of the differential plot where one would expect Monte Carlo error to dominate.

As highlighted above, when you do not have a test set for comparison, it may be hard to validate the optimal number of training points required for a good approximation. While at NLO we present the results of networks trained on 100k points, and found this number to be relatively optimal with regard to accuracy, stability and training time, we do not claim that this will always be the case for other processes. Although generating more \njet~matrix elements for testing is the best way to assess network accuracy, a possible substitute would be to test on the training data. While this is not generally regarded as good practice, given the problem at hand it may not be as bad as in other cases. For instance, unless there is a large degree of noise in the cross-section given the size of the training dataset, as an initial measure of model performance we can quantify the uncertainty in our training set and assess the proximity of our network uncertainties and this Monte Carlo error. Additionally, our network uncertainty calculation depends only on the network's behaviour relative to the training set and is independent of the test set. Therefore, although testing on the training set is still not ideal, given how we calculate our network uncertainties and by using our physics knowledge of the Monte Carlo error, we are able to use this as a first test of network performance without having to generate additional testing data.

\section{Conclusions \label{sec:conclusions}}

In this article we have explored the possibility of optimising simulations for
many scale processes needed for LHC analyses. Machine learning technology is
finding an increasing number of applications in particle physics and offers
the potential to dramatically reduce the CPU cost of expensive simulations.

The application to scattering amplitudes is a little different to classic examples of neural networks in
that the dataset is exact.\footnote{Technically we restrict to double precision, although higher precision arithmetic could be used in principle.} We can also have complete control over the range
of the dataset, although the CPU cost of obtaining the data can be very high. The challenge is to
make a sufficiently good fit to the data that a reliable interpolation and extrapolation of
differential cross-sections can be made. The CPU cost of the extrapolation/interpolation is
negligible in this procedure so the further the network can be extrapolated the better the
computational speed up.

In this study we have looked at multi-scale amplitudes which are not well suited to more
traditional approximations with polynomial grids. At one-loop scattering for $2\to 4$ or higher
multiplicity becomes extremely expensive even with modern automated tools. We find that a reliable
amplitude approximation can be difficult to achieve when using a single neural network due to the large
changes in the amplitude related to its singularity structure. We compare this naive approach to a
technique in which an ensemble of networks are use to approximate the amplitude by separating the
singularities using an NLO FKS partitioning.

Understanding the reliability of this approach is one of the biggest challenges. By varying the
initial data and parameter initialisations used in the network, we find a way to estimate the error
on the networks. For all but the highest multiplicity, $e^+e^- \to 5$ jets, we also provide
comparisons to direct integration of the amplitude. At LO we observe that the FKS partitioning
provides significantly more reliable and accurate estimates than the single network approach, while
in the case of NLO k-factors, where the leading order singularity structure is divided out, the
partitioning still helps in these regards with results accurate to within a few \%. Moreover, we show in Equations (\ref{FKS_normal i_0} - \ref{zeta p,s comparison}) that each network in our ensemble has a smaller associated uncertainty than that of the single network approach, thus suggesting that the ensemble model is learning the divergent structure with a higher confidence than the naive model. Indeed, this is the case at both LO and NLO. The networks not
only provide good scattering amplitude approximations, but also lead to reliable predictions with
at least a factor of 10 improvement to the complete simulation.

In this initial study we have made a number of simplifications whose effect could be important
when using the technique for a realistic analysis. Firstly, we employed a simple flat phase-space
generation using the RAMBO algorithm. This makes it hard to compare with the more efficient
generators used in state-of-the-art Monte-Carlo simulations. The JADE jet algorithm may exacerbate the soft
singularities and so the effect of alternative jet algorithms, as well as the effect of introducing initial
state singularities in $pp$ collisions, should be studied in the future. We also see in the higher
multiplicity cases that the error from the neural network approximation does start to increase.
It may be in these cases that the NLO FKS separation requires modification. In this study we used a
simple version of the partition function based only on the kinematic invariants. In general we can
alter the scaling power of the invariants in the various limits which will affect the behaviour
of the FKS regions away from the singularities. We may also find that effects of higher order, double
unresolved singularities begin to play a role. Since NNLO sector decomposition strategies are
available it would be interesting to explore this direction in the future. Another important step
would be to apply the technique to integration of infrared subtracted, real radiation. This case
is currently the most CPU expensive part of producing precise differential distributions for
comparison with the experiments.

\section{Acknowledgements \label{sec:acknowledgements}}

We are very grateful to Frank Krauss, Michael Spannowsky and Daniel M\^{a}itre for useful discussions. JB is supported by the UK Science and Technology Facilities Council (STFC) grant number ST/P006744/1.
SB is supported by an STFC Rutherford Fellowship ST/L004925/1. This project has received funding from the European Union’s Horizon
2020 research and innovation programme under grant agreement No 772099.

\bibliographystyle{JHEP}
\bibliography{nnamps_eejets.bib}

\providecommand{\href}[2]{#2}\begingroup\raggedright\begin{thebibliography}{10}

\bibitem{Czakon:2008zk}
M.~Czakon, \emph{{Tops from Light Quarks: Full Mass Dependence at Two-Loops in
  QCD}}, \href{http://dx.doi.org/10.1016/j.physletb.2008.05.028}{\emph{Phys.
  Lett.} {\bf B664} (2008) 307--314},
  [\href{http://arxiv.org/abs/0803.1400}{{\tt 0803.1400}}].

\bibitem{Borowka:2016ehy}
S.~Borowka, N.~Greiner, G.~Heinrich, S.~P. Jones, M.~Kerner, J.~Schlenk et~al.,
  \emph{{Higgs Boson Pair Production in Gluon Fusion at Next-to-Leading Order
  with Full Top-Quark Mass Dependence}},
  \href{http://dx.doi.org/10.1103/PhysRevLett.117.079901,
  10.1103/PhysRevLett.117.012001}{\emph{Phys. Rev. Lett.} {\bf 117} (2016)
  012001}, [\href{http://arxiv.org/abs/1604.06447}{{\tt 1604.06447}}].

\bibitem{Heinrich:2017kxx}
G.~Heinrich, S.~P. Jones, M.~Kerner, G.~Luisoni and E.~Vryonidou, \emph{{NLO
  predictions for Higgs boson pair production with full top quark mass
  dependence matched to parton showers}},
  \href{http://dx.doi.org/10.1007/JHEP08(2017)088}{\emph{JHEP} {\bf 08} (2017)
  088}, [\href{http://arxiv.org/abs/1703.09252}{{\tt 1703.09252}}].

\bibitem{Jones:2018hbb}
S.~P. Jones, M.~Kerner and G.~Luisoni, \emph{{Next-to-Leading-Order QCD
  Corrections to Higgs Boson Plus Jet Production with Full Top-Quark Mass
  Dependence}},
  \href{http://dx.doi.org/10.1103/PhysRevLett.120.162001}{\emph{Phys. Rev.
  Lett.} {\bf 120} (2018) 162001}, [\href{http://arxiv.org/abs/1802.00349}{{\tt
  1802.00349}}].

\bibitem{Heinrich:2019bkc}
G.~Heinrich, S.~P. Jones, M.~Kerner, G.~Luisoni and L.~Scyboz, \emph{{Probing
  the trilinear Higgs boson coupling in di-Higgs production at NLO QCD
  including parton shower effects}},
  \href{http://dx.doi.org/10.1007/JHEP06(2019)066}{\emph{JHEP} {\bf 06} (2019)
  066}, [\href{http://arxiv.org/abs/1903.08137}{{\tt 1903.08137}}].

\bibitem{Cybenko1989}
G.~Cybenko, \emph{Approximation by superpositions of a sigmoidal function},
  \href{http://dx.doi.org/10.1007/BF02551274}{\emph{Mathematics of Control,
  Signals and Systems} {\bf 2} (Dec, 1989) 303--314}.

\bibitem{Badger:2012pg}
S.~Badger, B.~Biedermann, P.~Uwer and V.~Yundin, \emph{{Numerical evaluation of
  virtual corrections to multi-jet production in massless QCD}},
  \href{http://dx.doi.org/10.1016/j.cpc.2013.03.018}{\emph{Comput. Phys.
  Commun.} {\bf 184} (2013) 1981--1998},
  [\href{http://arxiv.org/abs/1209.0100}{{\tt 1209.0100}}].

\bibitem{Bendavid:2017zhk}
J.~Bendavid, \emph{{Efficient Monte Carlo Integration Using Boosted Decision
  Trees and Generative Deep Neural Networks}},
  \href{http://arxiv.org/abs/1707.00028}{{\tt 1707.00028}}.

\bibitem{Klimek:2018mza}
M.~D. Klimek and M.~Perelstein, \emph{{Neural Network-Based Approach to Phase
  Space Integration}},  \href{http://arxiv.org/abs/1810.11509}{{\tt
  1810.11509}}.

\bibitem{Bothmann:2020ywa}
E.~Bothmann, T.~Janßen, M.~Knobbe, T.~Schmale and S.~Schumann,
  \emph{{Exploring phase space with Neural Importance Sampling}},
  \href{http://arxiv.org/abs/2001.05478}{{\tt 2001.05478}}.

\bibitem{Gao:2020zvv}
C.~Gao, S.~Höche, J.~Isaacson, C.~Krause and H.~Schulz, \emph{{Event
  Generation with Normalizing Flows}},
  \href{http://arxiv.org/abs/2001.10028}{{\tt 2001.10028}}.

\bibitem{Dinh2014NICENI}
L.~Dinh, D.~Krueger and Y.~Bengio, \emph{{NICE}: Non-linear independent
  components estimation},  in \emph{3rd International Conference on Learning
  Representations}, ICLR, 2015.

\bibitem{Otten:2018kum}
S.~Otten, K.~Rolbiecki, S.~Caron, J.-S. Kim, R.~Ruiz De~Austri and
  J.~Tattersall, \emph{{DeepXS: Fast approximation of MSSM electroweak cross
  sections at NLO}},
  \href{http://dx.doi.org/10.1140/epjc/s10052-019-7562-1}{\emph{Eur. Phys. J.}
  {\bf C80} (2020) 12}, [\href{http://arxiv.org/abs/1810.08312}{{\tt
  1810.08312}}].

\bibitem{NIPS2014_5423}
I.~Goodfellow, J.~Pouget-Abadie, M.~Mirza, B.~Xu, D.~Warde-Farley, S.~Ozair
  et~al., \emph{Generative adversarial nets},  in \emph{Advances in Neural
  Information Processing Systems 27} (Z.~Ghahramani, M.~Welling, C.~Cortes,
  N.~D. Lawrence and K.~Q. Weinberger, eds.), pp.~2672--2680.
\newblock Curran Associates, Inc., 2014.

\bibitem{Otten:2019hhl}
S.~Otten, S.~Caron, W.~de~Swart, M.~van Beekveld, L.~Hendriks, C.~van Leeuwen
  et~al., \emph{{Event Generation and Statistical Sampling for Physics with
  Deep Generative Models and a Density Information Buffer}},
  \href{http://arxiv.org/abs/1901.00875}{{\tt 1901.00875}}.

\bibitem{Hashemi:2019fkn}
B.~Hashemi, N.~Amin, K.~Datta, D.~Olivito and M.~Pierini, \emph{{LHC
  analysis-specific datasets with Generative Adversarial Networks}},
  \href{http://arxiv.org/abs/1901.05282}{{\tt 1901.05282}}.

\bibitem{DiSipio:2019imz}
R.~Di~Sipio, M.~Faucci~Giannelli, S.~Ketabchi~Haghighat and S.~Palazzo,
  \emph{{DijetGAN: A Generative-Adversarial Network Approach for the Simulation
  of QCD Dijet Events at the LHC}},
  \href{http://dx.doi.org/10.1007/JHEP08(2019)110}{\emph{JHEP} {\bf 08} (2020)
  110}, [\href{http://arxiv.org/abs/1903.02433}{{\tt 1903.02433}}].

\bibitem{Butter:2019cae}
A.~Butter, T.~Plehn and R.~Winterhalder, \emph{{How to GAN LHC Events}},
  \href{http://dx.doi.org/10.21468/SciPostPhys.7.6.075}{\emph{SciPost Phys.}
  {\bf 7} (2019) 075}, [\href{http://arxiv.org/abs/1907.03764}{{\tt
  1907.03764}}].

\bibitem{Butter:2019eyo}
A.~Butter, T.~Plehn and R.~Winterhalder, \emph{{How to GAN Event Subtraction}},
   \href{http://arxiv.org/abs/1912.08824}{{\tt 1912.08824}}.

\bibitem{Carrazza:2019cnt}
S.~Carrazza and F.~A. Dreyer, \emph{{Lund jet images from generative and
  cycle-consistent adversarial networks}},
  \href{http://dx.doi.org/10.1140/epjc/s10052-019-7501-1}{\emph{Eur. Phys. J.}
  {\bf C79} (2019) 979}, [\href{http://arxiv.org/abs/1909.01359}{{\tt
  1909.01359}}].

\bibitem{SHiP:2019gcl}
{\scshape SHiP} collaboration, C.~Ahdida et~al., \emph{{Fast simulation of
  muons produced at the SHiP experiment using Generative Adversarial
  Networks}},
  \href{http://dx.doi.org/10.1088/1748-0221/14/11/P11028}{\emph{JINST} {\bf 14}
  (2019) P11028}, [\href{http://arxiv.org/abs/1909.04451}{{\tt 1909.04451}}].

\bibitem{Bishara:2019iwh}
F.~Bishara and M.~Montull, \emph{{(Machine) Learning amplitudes for faster
  event generation}},  \href{http://arxiv.org/abs/1912.11055}{{\tt
  1912.11055}}.

\bibitem{github}
J.~Bullock, ``n3jet.'' \url{https://github.com/JosephPB/n3jet}, 2020.

\bibitem{Ossola:2006us}
G.~Ossola, C.~G. Papadopoulos and R.~Pittau, \emph{{Reducing full one-loop
  amplitudes to scalar integrals at the integrand level}},
  \href{http://dx.doi.org/10.1016/j.nuclphysb.2006.11.012}{\emph{Nucl. Phys.}
  {\bf B763} (2007) 147--169}, [\href{http://arxiv.org/abs/hep-ph/0609007}{{\tt
  hep-ph/0609007}}].

\bibitem{Bern:1994cg}
Z.~Bern, L.~J. Dixon, D.~C. Dunbar and D.~A. Kosower, \emph{{Fusing gauge
  theory tree amplitudes into loop amplitudes}},
  \href{http://dx.doi.org/10.1016/0550-3213(94)00488-Z}{\emph{Nucl. Phys.} {\bf
  B435} (1995) 59--101}, [\href{http://arxiv.org/abs/hep-ph/9409265}{{\tt
  hep-ph/9409265}}].

\bibitem{Britto:2004nc}
R.~Britto, F.~Cachazo and B.~Feng, \emph{{Generalized unitarity and one-loop
  amplitudes in N=4 super-Yang-Mills}},
  \href{http://dx.doi.org/10.1016/j.nuclphysb.2005.07.014}{\emph{Nucl. Phys.}
  {\bf B725} (2005) 275--305}, [\href{http://arxiv.org/abs/hep-th/0412103}{{\tt
  hep-th/0412103}}].

\bibitem{Ellis:2007br}
R.~K. Ellis, W.~T. Giele and Z.~Kunszt, \emph{{A Numerical Unitarity Formalism
  for Evaluating One-Loop Amplitudes}},
  \href{http://dx.doi.org/10.1088/1126-6708/2008/03/003}{\emph{JHEP} {\bf 03}
  (2008) 003}, [\href{http://arxiv.org/abs/0708.2398}{{\tt 0708.2398}}].

\bibitem{Giele:2008ve}
W.~T. Giele, Z.~Kunszt and K.~Melnikov, \emph{{Full one-loop amplitudes from
  tree amplitudes}},
  \href{http://dx.doi.org/10.1088/1126-6708/2008/04/049}{\emph{JHEP} {\bf 04}
  (2008) 049}, [\href{http://arxiv.org/abs/0801.2237}{{\tt 0801.2237}}].

\bibitem{Forde:2007mi}
D.~Forde, \emph{{Direct extraction of one-loop integral coefficients}},
  \href{http://dx.doi.org/10.1103/PhysRevD.75.125019}{\emph{Phys. Rev.} {\bf
  D75} (2007) 125019}, [\href{http://arxiv.org/abs/0704.1835}{{\tt
  0704.1835}}].

\bibitem{Berger:2008sj}
C.~F. Berger, Z.~Bern, L.~J. Dixon, F.~Febres~Cordero, D.~Forde, H.~Ita et~al.,
  \emph{{An Automated Implementation of On-Shell Methods for One-Loop
  Amplitudes}}, \href{http://dx.doi.org/10.1103/PhysRevD.78.036003}{\emph{Phys.
  Rev.} {\bf D78} (2008) 036003}, [\href{http://arxiv.org/abs/0803.4180}{{\tt
  0803.4180}}].

\bibitem{Badger:2008cm}
S.~D. Badger, \emph{{Direct Extraction Of One Loop Rational Terms}},
  \href{http://dx.doi.org/10.1088/1126-6708/2009/01/049}{\emph{JHEP} {\bf 01}
  (2009) 049}, [\href{http://arxiv.org/abs/0806.4600}{{\tt 0806.4600}}].

\bibitem{Berends:1987me}
F.~A. Berends and W.~T. Giele, \emph{{Recursive Calculations for Processes with
  n Gluons}}, \href{http://dx.doi.org/10.1016/0550-3213(88)90442-7}{\emph{Nucl.
  Phys.} {\bf B306} (1988) 759--808}.

\bibitem{Binoth:2010xt}
T.~Binoth et~al., \emph{{A Proposal for a Standard Interface between Monte
  Carlo Tools and One-Loop Programs}},
  \href{http://dx.doi.org/10.1016/j.cpc.2010.05.016}{\emph{Comput. Phys.
  Commun.} {\bf 181} (2010) 1612--1622},
  [\href{http://arxiv.org/abs/1001.1307}{{\tt 1001.1307}}].

\bibitem{Frixione:1995ms}
S.~Frixione, Z.~Kunszt and A.~Signer, \emph{{Three jet cross-sections to
  next-to-leading order}},
  \href{http://dx.doi.org/10.1016/0550-3213(96)00110-1}{\emph{Nucl. Phys.} {\bf
  B467} (1996) 399--442}, [\href{http://arxiv.org/abs/hep-ph/9512328}{{\tt
  hep-ph/9512328}}].

\bibitem{Frederix:2009yq}
R.~Frederix, S.~Frixione, F.~Maltoni and T.~Stelzer, \emph{{Automation of
  next-to-leading order computations in QCD: The FKS subtraction}},
  \href{http://dx.doi.org/10.1088/1126-6708/2009/10/003}{\emph{JHEP} {\bf 10}
  (2009) 003}, [\href{http://arxiv.org/abs/0908.4272}{{\tt 0908.4272}}].

\bibitem{Czakon:2014oma}
M.~Czakon and D.~Heymes, \emph{{Four-dimensional formulation of the
  sector-improved residue subtraction scheme}},
  \href{http://dx.doi.org/10.1016/j.nuclphysb.2014.11.006}{\emph{Nucl. Phys.}
  {\bf B890} (2014) 152--227}, [\href{http://arxiv.org/abs/1408.2500}{{\tt
  1408.2500}}].

\bibitem{Bartel:1986ua}
{\scshape JADE} collaboration, W.~Bartel et~al., \emph{{Experimental Studies on
  Multi-Jet Production in e+ e- Annihilation at PETRA Energies}},
  \href{http://dx.doi.org/10.1007/BF01410449}{\emph{Z. Phys.} {\bf C33} (1986)
  23}.

\bibitem{Kleiss:1985gy}
R.~Kleiss, W.~J. Stirling and S.~D. Ellis, \emph{{A New Monte Carlo Treatment
  of Multiparticle Phase Space at High-energies}},
  \href{http://dx.doi.org/10.1016/0010-4655(86)90119-0}{\emph{Comput. Phys.
  Commun.} {\bf 40} (1986) 359}.

\bibitem{Friedman}
F.~J. H. and W.~M. H., \emph{An adaptive importance sampling procedure},  1981.

\bibitem{Lepage:1977sw}
G.~P. Lepage, \emph{{A New Algorithm for Adaptive Multidimensional
  Integration}}, \href{http://dx.doi.org/10.1016/0021-9991(78)90004-9}{\emph{J.
  Comput. Phys.} {\bf 27} (1978) 192}.

\bibitem{Lepage:1980dq}
G.~P. Lepage, \emph{{VEGAS: An Adaptive Multidimensional Integration Program}},
   1980.

\bibitem{Ohl:1998jn}
T.~Ohl, \emph{{Vegas revisited: Adaptive Monte Carlo integration beyond
  factorization}},
  \href{http://dx.doi.org/10.1016/S0010-4655(99)00209-X}{\emph{Comput. Phys.
  Commun.} {\bf 120} (1999) 13--19},
  [\href{http://arxiv.org/abs/hep-ph/9806432}{{\tt hep-ph/9806432}}].

\bibitem{Kroeninger:2014bwa}
K.~Kroeninger, S.~Schumann and B.~Willenberg, \emph{{(MC)**3 -- a Multi-Channel
  Markov Chain Monte Carlo algorithm for phase-space sampling}},
  \href{http://dx.doi.org/10.1016/j.cpc.2014.08.024}{\emph{Comput. Phys.
  Commun.} {\bf 186} (2015) 1--10}, [\href{http://arxiv.org/abs/1404.4328}{{\tt
  1404.4328}}].

\bibitem{Press:1989vk}
W.~H. Press and G.~R. Farrar, \emph{{Recursive Stratified Sampling For
  Multidimensional Monte Carlo Iintegration}}, {\emph{Submitted to: Comp.in
  Phys.} (1989) }.

\bibitem{Jadach:2002kn}
S.~Jadach, \emph{{Foam: A General purpose cellular Monte Carlo event
  generator}},
  \href{http://dx.doi.org/10.1016/S0010-4655(02)00755-5}{\emph{Comput. Phys.
  Commun.} {\bf 152} (2003) 55--100},
  [\href{http://arxiv.org/abs/physics/0203033}{{\tt physics/0203033}}].

\bibitem{Draggiotis:2000gm}
P.~D. Draggiotis, A.~van Hameren and R.~Kleiss, \emph{{SARGE: An Algorithm for
  generating QCD antennas}},
  \href{http://dx.doi.org/10.1016/S0370-2693(00)00532-3}{\emph{Phys. Lett.}
  {\bf B483} (2000) 124--130}, [\href{http://arxiv.org/abs/hep-ph/0004047}{{\tt
  hep-ph/0004047}}].

\bibitem{vanHameren:2002tc}
A.~van Hameren and C.~G. Papadopoulos, \emph{{A Hierarchical phase space
  generator for QCD antenna structures}},
  \href{http://dx.doi.org/10.1007/s10052-002-1000-4}{\emph{Eur. Phys. J.} {\bf
  C25} (2002) 563--574}, [\href{http://arxiv.org/abs/hep-ph/0204055}{{\tt
  hep-ph/0204055}}].

\bibitem{Frederix:2010ne}
R.~Frederix, S.~Frixione, K.~Melnikov and G.~Zanderighi, \emph{{NLO QCD
  corrections to five-jet production at LEP and the extraction of
  $\alpha_s(M_Z)$}},
  \href{http://dx.doi.org/10.1007/JHEP11(2010)050}{\emph{JHEP} {\bf 11} (2010)
  050}, [\href{http://arxiv.org/abs/1008.5313}{{\tt 1008.5313}}].

\bibitem{keras}
F.~Chollet et~al., ``Keras.'' \url{https://github.com/fchollet/keras}, 2015.

\bibitem{tensorflow}
M.~Abadi, A.~Agarwal, P.~Barham, E.~Brevdo, Z.~Chen, C.~Citro et~al.,
  ``{TensorFlow}: Large-scale machine learning on heterogeneous systems.''
  \url{https://www.tensorflow.org/}, 2015.

\bibitem{adam}
D.~P. Kingma and J.~Ba, \emph{Adam: A method for stochastic optimization},
  {\emph{3rd International Conference for Learning Representations} (2015) },
  [\href{http://arxiv.org/abs/1412.6980}{{\tt 1412.6980}}].

\bibitem{goodfellow}
I.~Goodfellow, Y.~Bengio and A.~Courville, \emph{Deep Learning}.
\newblock MIT Press, 2016.

\bibitem{tagasovska:2018}
N.~Tagasovska and D.~Lopez-Paz, \emph{Single-model uncertainties for deep
  learning}, {\emph{NeurlPS} (2019) },
  [\href{http://arxiv.org/abs/1811.00908}{{\tt 1811.00908}}].

\bibitem{Gal:2016}
Y.~Gal, \emph{Uncertainty in Deep Learning}.
\newblock PhD thesis, University of Cambridge, 2016.

\bibitem{Nachman:2019dol}
B.~Nachman, \emph{{A guide for deploying Deep Learning in LHC searches: How to
  achieve optimality and account for uncertainty}},
  \href{http://arxiv.org/abs/1909.03081}{{\tt 1909.03081}}.

\bibitem{Nachman:2019yfl}
B.~Nachman and C.~Shimmin, \emph{{AI Safety for High Energy Physics}},
  \href{http://arxiv.org/abs/1910.08606}{{\tt 1910.08606}}.

\bibitem{Bollweg:2019skg}
S.~Bollweg, M.~Haußmann, G.~Kasieczka, M.~Luchmann, T.~Plehn and J.~Thompson,
  \emph{{Deep-Learning Jets with Uncertainties and More}},
  \href{http://arxiv.org/abs/1904.10004}{{\tt 1904.10004}}.

\bibitem{Englert:2018cfo}
C.~Englert, P.~Galler, P.~Harris and M.~Spannowsky, \emph{{Machine Learning
  Uncertainties with Adversarial Neural Networks}},
  \href{http://dx.doi.org/10.1140/epjc/s10052-018-6511-8}{\emph{Eur. Phys. J.}
  {\bf C79} (2019) 4}, [\href{http://arxiv.org/abs/1807.08763}{{\tt
  1807.08763}}].

\bibitem{Cranmer:2015bka}
K.~Cranmer, J.~Pavez and G.~Louppe, \emph{{Approximating Likelihood Ratios with
  Calibrated Discriminative Classifiers}},
  \href{http://arxiv.org/abs/1506.02169}{{\tt 1506.02169}}.

\bibitem{Frixione:2005vw}
S.~Frixione, E.~Laenen, P.~Motylinski and B.~R. Webber, \emph{{Single-top
  production in MC@NLO}},
  \href{http://dx.doi.org/10.1088/1126-6708/2006/03/092}{\emph{JHEP} {\bf 03}
  (2006) 092}, [\href{http://arxiv.org/abs/hep-ph/0512250}{{\tt
  hep-ph/0512250}}].

\end{thebibliography}\endgroup

\appendix
\section{Average tendencies of the mean squared error \label{AppendixA}}

This section is heavily borrowed from the appendices of \cite{Cranmer:2015bka, Nachman:2019dol} but is repeated here given the different applications of our specific problem.

For $d$-dimensional input data $X\in\mathbb{R}^{d}$ and targets $Y\in\mathbb{R}$, we train a network that acts as a function $f: \mathbb{R}^d \rightarrow \mathbb{R}$ by minimising a loss function, $L$, averaged over all points. In this domain, the distributions $X$ and $Y$ are clearly not independent and form a joint probability density $(X,Y)$. The output of the neural network, given some specific input variables, which minimises the loss function averaged over the entire training dataset is given by:

\begin{equation}\label{appendix_1}
l(x) = \text{argmin}_{f}(\EX[L(f(x), Y)|X=x]),
\end{equation}

where $\EX$ is the expectation value and $\text{argmin}_g(h(g(x)))$ denotes the values of a function $g$ that minimise $h$.

For the case of the mean squared error, $L(f(X), Y) = (f(X) - Y)^2$, Equation (\ref{appendix_1}) becomes:

\begin{align}
l(x) &= \text{argmin}_{f}(\EX[(f(x) - Y)^2|X=x])\\
&= \text{argmin}_{f}(\EX[f(x)^2 - 2f(x)Y + Y^2|X=x])\\
&= \text{argmin}_{f}(f(x)^2 - 2f(x)\EX[Y|X=x]).
\end{align}

Minimising $l(x)$ now gives: $l(x) = \EX[Y|X=x]$, thus demonstrating that a network, when trained using the mean squared error, approaches the average value of the target distribution.

\section{FKS pairs and partition functions \label{AppendixB}}

The FKS subtraction formalism was designed to provide a framework by which the divergent structure arising from the real radiation corrections at NLO can be constructed and subtracted in $(n+1)$ phase-space, where $n$ is the number of jets at the Born level, and added back in and solved analytically via dimensional regularisation \cite{Frixione:1995ms}:

\begin{equation}
\sigma_{\text{NLO}} = \int_n\text{d}\sigma^{(B)} + \int_m\left[\text{d}\sigma^{(V)} + \int_1 \text{d}\sigma^{(S)}\right] + \int_{n+1}[\text{d}\sigma^{(R)} - \text{d}\sigma^{(S)}],
\end{equation}

where $\sigma^{(B)}$ is the Born cross-section, $\sigma^{(R)}$ and $\sigma^{(V)}$ are the real and virtual corrections at NLO and $\sigma^{(S)}$ is the real singular structure. By performing subtraction we are able to ensure that the singular structures of the virtual and real corrections cancel, thus leaving us with a non-divergent NLO cross-section.

For the processes considered here, the most general way of defining FKS pairs is given by:

\begin{multline}\label{FKS pairs real}
\mathcal{P}_{\text{FKS}} = \{(i,j)\,|\,3 \leq\ i \leq n_g+2,\,\,3 \leq\ j \leq n_g+2, i \neq j, \\
\mathcal{M}^{(n+1,0)}\to\infty\,\, \text{if} \,\,p_i^0\to0\,\, \text{or} \,\,p_j^0\to0\,\, \text{or} \,\,\vec{p}_i||\vec{p}_j\},
\end{multline}

which is the equivalent definition as that used in Equation (\ref{FKS pairs}), but where for our purposes we have used the pairs defined by the Born and virtual correction divergent structures since we do not calculate real corrections and we are not trying to perform subtraction.

Given that FKS pairs are ordered, there is redundancy in Equation (\ref{FKS pairs real}) since we will double count the soft singularities. An alternative definition is just to drop the $p_j^0\to0$ criteria to get:

\begin{multline}\label{FKS pairs real redundant}
\mathcal{P}_{\text{FKS}} = \{(i,j)\,|\,3 \leq\ i \leq n_g+2,\,\,3 \leq\ j \leq n_g+2, i \neq j, \\
\mathcal{M}^{(n+1,0)}\to\infty\,\, \text{if} \,\,p_i^0\to0\,\, \text{or} \,\,\vec{p}_i||\vec{p}_j\},
\end{multline}

as shown in \cite{Frederix:2009yq}. By using the definition given in Equation (\ref{FKS pairs real redundant}), we end up with the general FKS criteria that \textit{each FKS partition contain at most one collinear and one soft singularity}. Formalising this mathematically allows us to require the following criteria be met by any such FKS partition function, $\mathcal{S}_{i,j}$ (adapted from \cite{Frederix:2009yq}):

\begin{equation}
\sum_{(i,j)\in\mathcal{P}_{\text{FKS}}}\mathcal{S}_{i,j} = 1,
\end{equation}
\begin{equation}
\lim_{\vec{p}_k||\vec{p}_l} \mathcal{S}_{i,j} = 0, \,\,\forall (k,l) \in \mathcal{P}_{\text{FKS}}\,\, \text{with}\,\, (k,l) \neq (i,j),
\end{equation}
\begin{equation}\label{S_i,j soft to zero}
\lim_{p_k^0\to0} \mathcal{S}_{i,j} = 0, \,\,\forall (k,l)\in \mathcal{P}_{\text{FKS}}\,\, \text{with}\,\, k \neq i.
\end{equation}

Examples of partition functions satisfying these conditions are given in \cite{Frederix:2009yq} in terms of energies and angles and in \cite{Frixione:2005vw} in terms of $s_{ij}$ variables among others.

While defining a function in terms of energies and angles can be beneficial when doing full FKS subtraction, for ease of computation we use the Lorentz invariant $s_{ij}$ variables defined in Equation (\ref{S_i,j definition}). However, we note that our definition of the FKS partition function does not satisfy Equation (\ref{S_i,j soft to zero}) and therefore some of our partitions will contain multiple soft singularities and thus result in redundancies.

\end{document}